\documentstyle[epsfig]{article}
\newcommand{\ba}{\begin{eqnarray}}
\newcommand{\ea}{\end{eqnarray}}
\begin{document}

\title{The Algebraic Cluster Model. Three-body clusters 
\footnote{Annals of Physics (N.Y.), in press}}
\author{R. Bijker \\
Dipartimento di Fisica, Universit\`a degli Studi di Genova, \\
Via Dodecaneso 33, I-16146 Genova, Italy 
\and
Instituto de Ciencias Nucleares,\\
Universidad Nacional Aut\'onoma de M\'exico,\\
A.P. 70-543, 04510 M\'{e}xico D.F., M\'{e}xico 
\footnote{Permanent address}
\and 
F. Iachello \\
Center for Thoretical Physics, Sloane Laboratory,\\
Yale University, New Haven, CT 06520-8120}
\date{March 26, 2002}
\maketitle

\begin{abstract}
A new method is introduced to study three-body clusters. Triangular
configurations with ${\cal D}_{3h}$ point-group symmetry are analyzed. The
spectrum, transition form factors and $B(E\lambda)$ values of $^{12}$C are
investigated. It is concluded that the low-lying spectrum of $^{12}$C can be
described by three alpha-particles at the vertices of an equilateral
triangle, but not as a rigid structure. Large rotation-vibration
interactions, Coriolis forces and vibration-vibration interactions are
needed. Other interpretations, such as the harmonic oscillator and 
a soft deformed oscillator with $SO(6)$ hyperspherical symmetry, 
appear to be excluded by electron scattering data.
\end{abstract}

\section{Introduction}

The Cluster Model of nuclei was introduced at the very beginning of Nuclear
Physics as a natural consequence of the molecular viewpoint in which
aggregates are formed by combining constituent units into composite objects.
In particular, the alpha-particle model represented, for many years, a
viable model of light nuclei \cite{wheeler}. With the successes of the
Nuclear Shell Model, the Cluster Model lost most of its appeal and the
interest shifted in `deriving' the alpha-particle model from the Nuclear
Shell Model \cite{wilder}.A large literature exists in this direction,
especially from the Japanese school \cite{ikeda,fujiwara,horiuchi}. In recent
years, clustering in nuclei has found renewed interest, since it appears
that many new isotopes of light nuclei far from stability possess a cluster
structure (not necessarely alpha). In view of this renewed interest, we have
readdressed the question of clustering in nuclei with two purposes in mind.
First, to analyze cluster configurations at the 'classical' level in
somewhat more detail than previously done, deriving explicit expressions for
the density distribution and its multipole moments, and second, and most
importantly, to introduce a formalism that allows one to study in a
relatively simple fashion the actual situation which occurs in nuclei, where
clustering is not a rigid-like molecular structure, but a rather soft
(liquid-like) structure. The new approach is based on the bosonic
quantization of many-body problems introduced by one of us \cite{iac}. In
this method, a quantum mechanical problem with $\nu$ space degrees of
freedom is quantized in terms of elements of the Lie algebra $U(\nu +1)$.
This method has been used in a variety of problems, ranging from Nuclear
Physics to Molecular Physics, from\ Hadronic Physics to\ Polymer Physics and
has proven to be very useful in analyzing experimental data. In a series of
articles beginning with the present one, we intend to exploit the algebraic
method for Cluster Physics. Two-body clusters, where the algebra is $U(4)$,
have been discussed previously \cite{iac1}. Here we begin with three-body
clusters, where the algebra is $U(7)$, and apply the method explicitly to
the study of three-alpha configurations in nuclei, in particular to 
$^{12}$C. Many authors have suggested that this nucleus in its ground state 
is composed of three alpha-particles at the vertices of an equilateral
triangle. In this article we study, both at the classical and at the 
quantum level,
a configuration of three-particles at the vertices of an equilateral
triangle (point-group symmetry ${\cal D}_{3h}$) and analyze the extent to
which the experimental data support this configuration. A preliminary
account of this part of our work has appeared \cite{BI,aurora}. 
In addition to the
study of properties of triangular configurations we also discuss other
possible arrangements of three particles and their motion, in particular
those corresponding to a six-dimensional spherical oscillator and to a
six-dimensional deformed oscillator. Altough these configurations do not
appear to describe well the experimental data in $^{12}$C the formulas we
derive may be of interest in other situations involving three particles.

\section{Classical treatment}

The treatment of a triangular configuration in classical mechanics is very
well known from molecular physics. However, this treatment is usualy done in
terms of point-like constituents. Although the generalization to composite
constituents is straighforward, in this introductory section we present some
new and more detailed results.

\subsection{Energies}

We consider three identical particles of mass $Am/3$ at the vertices of an
equilateral triangle (point-group symmetry ${\cal D}_{3h}$) (see 
Fig.~\ref{geometry}). The origin is placed at the center of mass 
and the three particles span the $xz$ plane. The distance from the center 
is $\beta$. The spherical coordinates of the three particles 
$\vec{r}_{i}=(r_{i},\theta_{i},\phi_{i})$ are then $(\beta,0,0)$,  
$(\beta,2\pi/3,\pi)$ and $(\beta,2\pi/3,0)$ for $i=1,2$ and $3$, respectively.
For a mass distribution $\rho (\vec{r})$, rotational energies can be
computed by evaluating the moments of inertia 
\begin{eqnarray}
{\cal I}_{x} &=& \int (y^{2}+z^{2}) \rho(\vec{r}) \, d\vec{r} ~,  
\nonumber \\
{\cal I}_{y} &=& \int (z^{2}+x^{2}) \rho(\vec{r}) \, d\vec{r} ~,  
\nonumber \\
{\cal I}_{z} &=& \int (x^{2}+y^{2}) \rho(\vec{r}) \, d\vec{r} ~.
\end{eqnarray}
The rotational energy is simply 
\begin{eqnarray}
E_{{\rm rot}} &=& \frac{1}{2{\cal I}_{x}} L_{x}^{2}
+ \frac{1}{2{\cal I}_{y}} L_{y}^{2} + \frac{1}{2{\cal I}_{z}} L_{z}^{2} ~.
\end{eqnarray}

\subsubsection{Point-like masses}

For a point-like mass distribution 
\begin{eqnarray}
\rho (\vec{r})&=&\frac{Am}{3}\sum_{i=1}^{3}\delta (\vec{r}-\vec{r}_{i})~,
\end{eqnarray}
a simple calculation gives 
\begin{eqnarray}
{\cal I}_{x} \;=\; {\cal I}_{z} &=& \frac{1}{2} Am \beta^{2}~,  
\nonumber \\
{\cal I}_{y} &=& Am \beta^{2} ~,
\end{eqnarray}
and thus 
\begin{eqnarray}
E_{{\rm rot}} &=& \frac{1}{Am\beta^2} 
\left[ \vec{L}^{2} - \frac{1}{2} K^{2} \right] ~,
\label{erot}
\end{eqnarray}
where $\vec{L}$ is the angular momentum and $K$ its projection on the
three-fold $y$-axis.

\subsubsection{Extended mass distribution}

Consider the case in which instead of point masses there are, at the
vertices of the triangle, extended distributions. Since we have in mind
applications to the alpha-particle model, we consider here the case in which
the distribution is Gaussian 
\begin{eqnarray}
\rho (\vec{r}) &=& \frac{Am}{3} \left( \frac{\alpha }{\pi }\right)^{3/2} 
\sum_{i=1}^{3}\exp \left[ -\alpha \left( \vec{r}-\vec{r}_{i}\right)^{2} 
\right] ~.
\end{eqnarray}
It is straightforward to compute the moments of inertia 
\begin{eqnarray}
{\cal I}_{x} \;=\; {\cal I}_{z} &=& \frac{1}{2} Am \beta^{2} 
\left( 1+\frac{2}{\alpha \beta^{2}} \right) ~,  
\nonumber \\
{\cal I}_{y} &=& Am \beta^{2} \left( 1+\frac{1}{\alpha \beta^{2}} \right) ~. 
\label{inertia}
\end{eqnarray}
The rotational energy is now given by 
\begin{eqnarray}
E_{{\rm rot}} &=& \frac{1}{Am \beta^{2} \left( 1+\frac{2}{\alpha \beta^{2}} 
\right)} \left[ \vec{L}^{2} - \frac{\alpha \beta^{2}} 
{2(1+\alpha \beta ^{2})}K^{2}\right] ~.
\end{eqnarray}
For typical values of $\alpha $ and $\beta $, as for example those extracted
from the form factors of $^{12}$C (discussed in the following sections), $%
\alpha =0.52$ fm$^{-2}$ and $\beta =1.74$ fm, we find that the coefficient
in front of $K^{2}$ reduces from $-0.50$ to $-0.30$. We note in general that
the rotational energy of a symmetric top is given by 
\begin{eqnarray}
E_{{\rm rot}} &=& A \, \vec{L}^{2} + B \, K^{2} ~.
\end{eqnarray}
The top is called oblate if $B<0$, prolate if $B>0$ and spherical if $B=0$.
Three particles on a plane are always oblate. The extended distribution
tends to make the top less oblate.

Vibrational energies can be computed by tying the particles with strings and
evaluating the eigenfrequencies \cite{herzberg}. Denoting by $q_{1}$, 
$q_{2}$ and $q_{3}$ the displacements of the sides of the triangle and 
assuming a potential energy of the type 
\begin{eqnarray}
V &=& \frac{1}{2}C\left( q_{1}^{2}+q_{2}^{2}+q_{3}^{2}\right) +D\left(
q_{1}q_{2}+q_{2}q_{3}+q_{3}q_{1}\right) ~,
\end{eqnarray}
one can obtain the frequencies of the normal modes as 
\begin{eqnarray}
\omega_{1} &=& \sqrt{\frac{9(C+2D)}{Am}} ~,  
\nonumber \\
\omega_{2} &=& \sqrt{\frac{9(C-D)}{2Am}} ~.
\end{eqnarray}
The vibration $\omega_{1}$ is singly degenerate, while the vibration 
$\omega_{2}$ is doubly degenerate.

\subsection{Transition probabilities}

Consider now the case in which the three particles have a charge $Ze/3$ 
(the case in which the particles have a magnetic moment will not be 
discussed here since we have in mind the alpha-particle model, where the 
constituents have no magnetic moment). The total charge distribution can 
be expanded into multipoles 
\begin{eqnarray}
\rho (\vec{r}) &=& \sum_{\ell m}A_{\ell m}(r) \, Y_{\ell m}(\theta,\phi) ~,
\end{eqnarray}
with 
\begin{eqnarray}
A_{\ell m}(r) &=& \int Y_{\ell m}^{\ast}(\theta,\phi) \, \rho(\vec{r}) 
\, d\cos\theta \, d\phi ~.
\end{eqnarray}
Instead of the charge distribution, one can consider the form factor, given
by its Fourier transform 
\begin{eqnarray}
F(\vec{q}) &=& \frac{1}{Ze} \int e^{i \vec{q} \cdot \vec{r}} \, 
\rho(\vec{r}) \, d\vec{r} \;=\; 
\sum_{\ell m} F_{\ell m}(q) \, Y_{\ell m}(\hat{q})~,
\end{eqnarray}
with 
\begin{eqnarray}
F_{\ell m}(q) &=& \frac{4\pi}{Ze} \int r^{2} \, A_{\ell m}(r) \, i^{l} \, 
j_{\ell}(qr) \, dr~.
\end{eqnarray}
The multipole moments of the charge configuration are 
\begin{eqnarray}
Q_{\ell m} &=& \int r^{l} \, Y_{\ell m}^{\ast}(\theta,\phi) \, 
\rho(\vec{r}) \, d\vec{r} \;=\; \int A_{\ell m}(r) \, r^{\ell +2} \, dr  
\nonumber \\
&=& \frac{Ze}{4\pi} \, (2\ell +1)!! \, \lim_{q \rightarrow 0} 
\frac{F_{\ell m}(q)}{i^{\ell} q^{\ell}} ~.
\end{eqnarray}
From these multipole moments, the transition probabilities can be calculated
classically \cite{jackson}. The transition probability per unit time is 
\begin{eqnarray}
T(E\ell) &=& 8\pi c \, \frac{e^{2}}{hc} \, \frac{\ell +1}{\ell \left[ 
(2\ell +1)!!\right]^{2}} \, k^{2\ell +1}\,B(E\ell)~,
\end{eqnarray}
where 
\begin{eqnarray}
B(E\ell) &=& \sum_{m=-\ell }^{\ell }Q_{\ell m}^{\ast }Q_{\ell m}~.
\end{eqnarray}

\subsubsection{Point charges}

For a point-like charge distribution 
\begin{eqnarray}
\rho (\vec{r}) &=& \frac{Ze}{3}\sum_{i=1}^{3}\delta (\vec{r}-\vec{r}_{i})~,
\label{charge}
\end{eqnarray}
the form factor is given by 
\begin{eqnarray}
F(\vec{q})&=&\frac{1}{3}\sum_{i=1}^{3}e^{i\vec{q}\cdot \vec{r}_{i}}~. 
\label{fpoint}
\end{eqnarray}
The corresponding $B(E\ell )$ values are 
\begin{eqnarray}
B(E\ell) &=& \left(\frac{Ze}{3}\right)^{2} \, \beta ^{2\ell} \, 
\frac{2\ell +1}{4\pi} \, \left[ 3+6P_{\ell}(-\frac{1}{2}) \right] ~,
\end{eqnarray}
which gives 
\begin{eqnarray}
B(E2) &=& (Ze)^{2} \, \frac{5}{4\pi} \, \frac{1}{4} \, \beta^{4} ~,  
\nonumber \\
B(E3) &=& (Ze)^{2} \, \frac{7}{4\pi} \, \frac{5}{8} \, \beta^{6} ~,  
\nonumber \\
B(E4) &=& (Ze)^{2} \, \frac{9}{4\pi} \, \frac{9}{64}\, \beta^{8} ~.  
\label{bel}
\end{eqnarray}
The dipole radiation $B(E1)$ vanishes because of the spatial symmetry of the 
charge distribution. 

\subsubsection{Extended charge distributions}

Next we consider the case in which instead of point charges there are, 
at the vertices of the triangle, extended distributions 
\begin{eqnarray}
\rho (\vec{r}) &=&\frac{Ze}{3}\left( \frac{\alpha }{\pi }\right)
^{3/2}\sum_{i=1}^{3}\exp \left[ -\alpha \left( \vec{r}-\vec{r}_{i}\right)
^{2}\right]  \nonumber \\
&=&\frac{Ze}{3}\left( \frac{\alpha }{\pi }\right) ^{3/2}e^{-\alpha
(r^{2}+\beta ^{2})}\,4\pi \,\sum_{i=1}^{3}\sum_{\lambda =0}^{\infty
}i_{\lambda }(2\alpha \beta r)\,Y_{\lambda }(\theta ,\phi )\cdot Y_{\lambda
}^{\ast }(\theta _{i},\phi _{i})~,  \label{rhor}
\end{eqnarray}
where $i_{\lambda }(x)=j_{\lambda }(ix)/i^{\lambda }$ is the modified
spherical Bessel function. The corresponding form factor is 
that of the point-like charge distribution multiplied by an exponential 
\begin{eqnarray}
F(\vec{q})&=&\frac{1}{3}\,e^{-q^{2}/4\alpha }\sum_{i=1}^{3}e^{i\vec{q}%
\cdot \vec{r}_{i}}~.  \label{fq}
\end{eqnarray}
The $B(E\ell)$ values of the extended charge distribution are the same as
those of the point-like configuration.

\section{Quantum treatment}

The quantum treatment of an identical three-body cluster can be done in
several ways. The most straightforward is to introduce relative Jacobi 
coordinates 
\begin{eqnarray}
\vec{\rho} &=& \left( \vec{r}_{1} - \vec{r}_{2} \right) /\sqrt{2} ~,  
\nonumber \\
\vec{\lambda} &=& \left( \vec{r}_{1} + \vec{r}_{2} - 2\vec{r}_{3} \right)/
\sqrt{6} ~,
\end{eqnarray}
write down a Hamiltonian in terms of these coordinates and their canonically
conjugate momenta $\vec{p}_{\rho }$ and $\vec{p}_{\lambda }$, and solve the
Schr\"odinger equation 
\begin{eqnarray}
\left[ \frac{1}{2m}(\vec{p}_{\rho }^{\,2}+\vec{p}_{\lambda }^{\,2}) 
+ V(\vec{\rho},\vec{\lambda}) \right] \psi (\vec{\rho},\vec{\lambda})
&=& E \, \psi (\vec{\rho},\vec{\lambda}) ~, 
\end{eqnarray}
to obtain the energy eigenvalues and eigenvectors. 

\subsection{Energies}

It is convenient to make a change of variables to coordinates ($r$, $\xi$, 
$\Omega_{\rho}$, $\Omega _{\lambda}$) which are related to the Jacobi
coordinates ($\rho$, $\Omega_{\rho}$, $\lambda$, $\Omega_{\lambda}$) by 
\begin{eqnarray}
\rho \;=\; r\sin \xi ~,&\hspace{1cm}& \lambda \;=\;r\cos \xi ~.
\end{eqnarray}
In the new coordinates, the kinetic energy is given by \cite{Smith,fabre}
\begin{eqnarray}
-\frac{1}{2mr^{5}}\frac{\partial }{\partial r}\left( r^{5}\frac{\partial }{%
\partial r}\right) +\frac{1}{2mr^{2}}\left[ -\frac{\partial ^{2}}{\partial
\xi ^{2}}-4\cot 2\xi \frac{\partial }{\partial \xi }+\frac{\vec{L}_{\rho
}^{2}(\Omega _{\rho })}{\sin ^{2}\xi }+\frac{\vec{L}_{\lambda }^{2}(\Omega
_{\lambda })}{\cos ^{2}\xi }\right] ~.
\end{eqnarray}
In this section, we consider three different potentials, namely the 
harmonic oscillator, the deformed oscillator and the oblate symmetric top. 
For potentials that only depend on the hyperradius $r$, the Schr\"odinger
equation can be solved by separation of variables into an angular and a
radial equation 
\begin{eqnarray}
\left[ - \frac{\partial^2}{\partial \xi^2} - 4 \cot 2\xi \frac{\partial}{%
\partial \xi} + \frac{\vec{L}_{\rho}^2(\Omega_{\rho})}{\sin^2 \xi} + \frac{%
\vec{L}_{\lambda}^2(\Omega_{\lambda})}{\cos^2 \xi} \right]
\psi(\xi,\Omega_{\rho},\Omega_{\lambda}) &=& \sigma(\sigma+4) \, 
\psi(\xi,\Omega_{\rho},\Omega_{\lambda}) ~,  
\nonumber \\
\left[ -\frac{1}{2m} \frac{1}{r^5} \frac{\partial}{\partial r} \left( r^5 
\frac{\partial}{\partial r} \right) + \frac{\sigma(\sigma+4)}{2mr^2} 
+ V(r) \right] \psi(r) &=& E \, \psi(r) ~.
\end{eqnarray}
For the six-dimensional harmonic oscillator 
\begin{eqnarray}
V(r) &=& \frac{1}{2}Cr^{2}~,
\end{eqnarray}
the energy spectrum can be obtained exactly 
\begin{eqnarray}
E(n) &=& \epsilon \, (n+3) ~, 
\label{ener1}  
\end{eqnarray}
with $n=0,1,\ldots$~, and $\epsilon =\sqrt{C/m}$. The allowed values of 
$\sigma$ are $\sigma=n,n-2,\ldots,1$ or $0$ for $n$ odd or even, 
respectively. 

For the six-dimensional displaced (or deformed) oscillator 
\begin{eqnarray}
V(r)&=&\frac{1}{2}C(r-r_{0})^{2}~,
\end{eqnarray}
the energy eigenvalues cannot be obtained exactly. An approximate solution,
valid in the limit of small oscillations around the equilibrium value 
$r_{0}$ is 
\begin{eqnarray}
E(v,\sigma) &\cong& \epsilon \, (v+\frac{1}{2})
+\frac{1}{2mr_{0}^{2}} \left[ \sigma (\sigma +4)+\frac{15}{4}\right] ~,
\label{ener2}
\end{eqnarray}
with $\epsilon=\sqrt{C/m}$. The first term gives a harmonic vibrational 
spectrum with $v=0,1,\ldots$~, and the second term gives the 
rotational spectrum with $\sigma =0,1,\ldots$~. 

In general, the potential is not invariant under six-dimensional rotations
as in the previous two examples, but only under rotations in three
dimensions. One can introduce the total angular momentum $\vec{I}=\vec{L}%
_{\rho }+\vec{L_{\lambda }}$, and make a transformation of variables to the
relative angle $2\theta $ between the two vectors 
\begin{eqnarray}
\cos 2\theta &=& \hat{\rho}\cdot \hat{\lambda}~,
\end{eqnarray}
and the Euler angles $\Omega $ of the intrinsic frame defined by 
\cite{palumbo} 
\begin{eqnarray}
\hat{1}\;=\;\frac{\hat{\rho}\times \hat{\lambda}}{\sin 2\theta }~,\hspace{1cm%
}\hat{2}\;=\;\frac{\hat{\rho}-\hat{\lambda}}{2\sin \theta }~,\hspace{1cm}%
\hat{3}\;=\;\frac{\hat{\rho}+\hat{\lambda}}{2\cos \theta }~.
\end{eqnarray}
In the new variables, the angular momentum dependent terms in the kinetic
energy are given by 
\begin{eqnarray}
\frac{\vec{L}_{\rho }^{2}(\Omega _{\rho })}{\sin ^{2}\xi }+\frac{\vec{L}%
_{\lambda }^{2}(\Omega _{\lambda })}{\cos ^{2}\xi } &=&\frac{1}{4\sin
^{2}\xi \cos ^{2}\xi }\left( \vec{I}^{\,2}+\tan ^{2}\theta \,I_{2}^{2}+\cot
^{2}\theta \,I_{3}^{2}-\frac{\partial ^{2}}{\partial \theta ^{2}}-2\cot
2\theta \frac{\partial }{\partial \theta }\right)  \nonumber \\
&&+\frac{\cos 2\xi }{2\sin ^{2}\xi \cos ^{2}\xi }\left( -\cot \theta
\,I_{2}I_{3}-\tan \theta \,I_{3}I_{2}+iI_{1}\frac{\partial }{\partial \theta 
}\right) ~.
\end{eqnarray}
The potential only depends on the intrinsic variables $r$, $\xi$ and $%
\theta$. An interesting situation occurs when the potential has sharp
minima in $r$, $\xi$ and $\theta$ 
\begin{eqnarray}
V(r,\xi,\theta) &=& \frac{1}{2} C (r-r_{0})^{2} 
+ \frac{1}{2} A (\xi-\xi_{0})^{2} 
+ \frac{1}{2} B (\theta-\theta_{0})^{2}~.
\end{eqnarray}
In the limit of small oscillations around $r_{0}$, $\xi_{0}=\pi/4$ and 
$\theta_{0}=\pi/4$, rotations and vibrations decouple, and
the set of resulting differential equations can be solved in closed form 
\begin{eqnarray}
E(v_{1},v_{2a},v_{2b},I,K) &\cong& \epsilon_{1} \, (v_{1}+\frac{1}{2}) 
+ \epsilon_{2a} \, (v_{2a}+\frac{1}{2}) 
+ \epsilon_{2b} \, (v_{2b}+\frac{1}{2})  \nonumber \\
&& + \frac{1}{mr_{0}^{2}} \left[ I(I+1) - \frac{1}{2}K^{2} 
-\frac{9}{8} \right] ~,  \label{ener3} 
\end{eqnarray}
with $\epsilon_{1}=\sqrt{C/m}$, $\epsilon_{2a}=\sqrt{A/mr_{0}^{2}}$ and $%
\epsilon_{2b}=\sqrt{B/mr_{0}^{2}}$. Here $K$ is the projection of the
angular momentum $I$ on the symmetry axis $\hat{1}$.

For three identical masses, the potential has to be invariant under their
permutation, i.e. the coefficients $A$ and $B$ are equal. 
As a consequence, the 
vibrations $v_{2a}$ and $v_{2b}$ form a doubly degenerate vibration, and it
is convenient to use instead of $v_{2a}$ and $v_{2b}$ the quantum numbers 
$v_{2}^{\ell _{2}}$ where $\ell _{2}=v_{2},v_{2}-2,\ldots,1$ or $0$ for $%
v_{2}=v_{2a}+v_{2b}$ odd or even. The vibrational energies can then be
written as 
\begin{eqnarray}
E_{{\rm vib}}(v_{1},v_{2}) &=& \epsilon _{1} \, (v_{1}+\frac{1}{2}) 
+ \epsilon_{2} \, (v_{2}+1) ~.
\label{vener3}
\end{eqnarray}
The rotational energies
\begin{eqnarray}
E_{{\rm rot}}(I,K) &=& \frac{1}{mr_{0}^{2}} 
\left[ I(I+1) - \frac{1}{2}K^{2} \right] ~,
\label{rener3}
\end{eqnarray}
are the same as derived classically for a triangular distribution of three
identical point-like masses of Eq.~(\ref{erot}) with $\vec{L}^{2}$ replaced 
by $I(I+1)$. The spectrum is that of an oblate symmetric top.

The triangular configuration with three identical masses has discrete
symmetry ${\cal D}_{3h}$. Among the quantum numbers, it is convenient to add
the parity $P$ and the transformation properties of the states under ${\cal D%
}_{3h}$. Since ${\cal D}_{3h}\sim {\cal D}_{3}\times P$, the transformation
properties under ${\cal D}_{3h}$ are labeled by parity and the
representations of ${\cal D}_{3}\sim S_{3}$. ${\cal D}_{3}$ has two
one-dimensional representations, usually denoted by $A_{1}$ and $A_{2}$, and
a two-dimensional representation denoted by $E$. The corresponding Young
tableau of the permutation group $S_{3}$ are 
\begin{eqnarray}
t=A_{1} &:& 
\begin{array}{l}
\setlength{\unitlength}{1.0pt}\begin{picture}(30,10)(0,0) \thinlines \put (
0, 0) {\line (1,0){30}} \put ( 0,10) {\line (1,0){30}} \put ( 0, 0) {\line
(0,1){10}} \put (10, 0) {\line (0,1){10}} \put (20, 0) {\line (0,1){10}}
\put (30, 0) {\line (0,1){10}} \end{picture}
\end{array}
\\
&&  \nonumber \\
t=E &:& 
\begin{array}{l}
\setlength{\unitlength}{1.0pt}\begin{picture}(20,20)(0,5) \thinlines \put (
0, 5) {\line (1,0){10}} \put ( 0,15) {\line (1,0){20}} \put ( 0,25) {\line
(1,0){20}} \put ( 0, 5) {\line (0,1){20}} \put (10, 5) {\line (0,1){20}}
\put (20,15) {\line (0,1){10}} \end{picture}
\end{array}
\\
&&  \nonumber \\
t=A_{2} &:& 
\begin{array}{l}
\setlength{\unitlength}{1.0pt}\begin{picture}(10,30)(0,10) \thinlines \put (
0, 10) {\line (1,0){10}} \put ( 0, 20) {\line (1,0){10}} \put ( 0, 30)
{\line (1,0){10}} \put ( 0, 40) {\line (1,0){10}} \put ( 0, 10) {\line
(0,1){30}} \put (10, 10) {\line (0,1){30}} \end{picture}
\end{array}
\end{eqnarray}
For three identical particles, the wave functions must transform as the 
symmetric representations $A_{1}$ of ${\cal D}_{3} \sim S_{3}$. 

\subsection{Transition probabilities}

Transition probabilities, charge radii and other electromagnetic properties  
of interest can be obtained from the transition form factors. For electric 
transitions (the only ones discussed here) the form factors are the matrix 
elements of the Fourier transform of the charge distribution 
\ba
F(i \rightarrow f;\vec{q}) &=& \int d\vec{r} \, 
e^{i\vec{q} \cdot \vec{r}} \, \langle \gamma_f,L_f,M_f \, 
| \, \hat{\rho}(\vec{r}) \, | \, \gamma_i,L_i,M_i \rangle ~, 
\ea
where $\gamma$ denotes the quantum numbers needed to classify the states 
uniquely. For the point-like charge configuration of Eq.~(\ref{charge}), 
the transition form factor reduces to 
\ba
F(i \rightarrow f;\vec{q}) &=& Ze \sum_M {\cal D}^{(L_f)}_{M_fM}(\hat{q}) 
\, {\cal F}(i \rightarrow f;q) \, {\cal D}^{(L_i)}_{MM_i}(-\hat{q}) ~, 
\ea
with
\ba
{\cal F}(i \rightarrow f;q) &=& \langle \gamma_f,L_f,M \, 
| \, e^{-i q \sqrt{2/3} \lambda_{z}} \, | \, \gamma_i,L_i,M \rangle ~. 
\label{calff}
\ea 
Here we have used the permutation symmetry of the initial and final 
wave functions, a transformation to Jacobi coordinates, and an 
integration over the center-of-mass coordinate. 
In Table~\ref{ff} we show some form factors for the transition from the 
ground state to an excited state ${\cal F}(0^+_1 \rightarrow L^P_i;q)$. 
The form factors for harmonic oscillator show an exponential dependence on 
the momentum transfer \cite{BIL,hoqm}, whereas for the deformed oscillator and 
the oblate top they behave as cylindrical \cite{BL2} and spherical Bessel 
functions \cite{BI,BIL,Inopin}, respectively. 
The form factors associated with the extended charge distribution of 
Eq.~(\ref{rhor}) can be obtained, as in the classical case, by multiplying 
the results from Table~\ref{ff} by a Gaussian $\exp(-q^2/4\alpha)$, 
which represents the form factor of the $\alpha$ particle. 

The transition probability $B(EL)$ can be extracted from the form factors  
in the long wavelength limit 
\ba
B(EL;i \rightarrow f) &=& \frac{[(2L+1)!!]^{2}}{4\pi (2L_{i}+1)} 
\, \lim_{q\rightarrow 0} \sum_{M_{f}M_{i}} 
\left| F(i \rightarrow f;\vec{q}) \right|^{2}/q^{2L} 
\nonumber\\
&=& (Ze)^2 \, \frac{[(2L+1)!!]^{2}}{4\pi (2L_{i}+1)} 
\, \lim_{q\rightarrow 0} \sum_M 
\left| {\cal F}(i \rightarrow f;q) \right|^{2}/q^{2L} ~. 
\label{belif}
\ea
In the case of the oblate symmetric top, we find for the rotational 
excitations of the ground state the same results as for the multipole 
radiation in the classical case (see Eq.~(\ref{bel})) 
\ba
B(E2;0^+ \rightarrow 2^+) &=& (Ze)^{2} \frac{5}{4\pi} \frac{1}{4} \beta^{4} ~,
\nonumber \\
B(E3;0^+ \rightarrow 3^-) &=& (Ze)^{2} \frac{7}{4\pi} \frac{5}{8} \beta^{6} ~,
\nonumber \\
B(E4;0^+ \rightarrow 4^+) &=& (Ze)^{2} \frac{9}{4\pi} \frac{9}{64}\beta^{8} ~.
\ea
The ground state rotational band of the oblate top does not have a 
symmetric state with $L^P=1^-$, which is in agreement with the absence 
of dipole radiation in the classical treatment. 

The ground state charge distribution can be extracted from the elastic 
form factor by taking the Fourier transform
\ba
\rho(r) &=& \frac{1}{(2\pi)^{3}} \int d\vec{q} \, 
F(0_{1}^{+} \rightarrow 0_{1}^{+};\vec{q}) \, e^{-i\vec{q}\cdot \vec{r}} 
\nonumber\\
&=& \frac{Ze}{2\pi^{2}} \int q^2 dq \, j_0(qr) \,   
{\cal F}(0_{1}^{+} \rightarrow 0_{1}^{+};q) ~. 
\label{gsch}
\ea
For the case of the oblate top, the ground state charge distribution 
can be derived as 
\ba
\rho(r) &=& Ze \left( \frac{\alpha }{\pi} \right)^{3/2} \, 
\frac{1}{4\alpha \beta r} 
\, \left[ e^{-\alpha (\beta -r)^{2}}-e^{-\alpha (\beta +r)^{2}}\right] ~.
\label{otch}
\ea
The charge radius can be obtained from the slope of the elastic form 
factor in the origin 
\ba
\langle r^{2} \rangle^{1/2} &=& \left[ -6 \left. 
\frac{d {\cal F}(0_{1}^{+} \rightarrow 0_{1}^{+};q)}{dq^{2}} \right|_{q=0} 
\right]^{1/2} ~. 
\ea
In Table~\ref{ff} we have introduced a scale parameter $\beta$ in the 
form factors, such that the charge radius for the harmonic oscillator, 
the deformed oscillator and the oblate top is given by  
\ba
\langle r^{2} \rangle^{1/2} &=& \sqrt{\frac{3}{2\alpha}+\beta^{2}} ~. 
\ea

\section{Algebraic treatment}

In this section, we discuss an algebraic treatment of three-cluster 
systems, in which the eigenvalue problem is solved by matrix diagonalization 
instead of by solving a set of differential equations. In the algebraic 
cluster model (ACM), the method of bosonic quantization is used. This
method consists in quantizing the Jacobi coordinates and momenta with boson
operators and adding an additional scalar boson \cite{BIL}
\ba
b_{\rho,m}^{\dagger} \;,\; b_{\lambda,m}^{\dagger} \;,\; s^{\dagger}
&\equiv& c_{\alpha}^{\dagger} \hspace{1cm} (m=0,\pm 1) 
\hspace{1cm} (\alpha=1,\ldots,7)  
\ea
The role of the additional scalar boson is to compactify the space thus making
calculations of matrix elements simpler. The set of 49 bilinear products of
creation and annihilation operators spans the Lie algebra of $U(7)$ 
\begin{eqnarray}
{\cal G} &:& G_{\alpha \beta} \;=\; c_{\alpha}^{\dagger} c_{\beta} 
\hspace{1cm} (\alpha,\beta=1,\ldots,7)
\end{eqnarray}
All operators are expanded into elements of this algebra.

The most generic Hamiltonian which includes up to two-body terms and is
invariant under the permutation group $S_{3}$, contains 10 parameters that
describe the relative motion of a system of three identical clusters. It can
be written as \cite{BIL} 
\begin{eqnarray}
H &=&\epsilon _{0}\,s^{\dagger }\tilde{s}-\epsilon _{1}\,(b_{\rho }^{\dagger
}\cdot \tilde{b}_{\rho }+b_{\lambda }^{\dagger }\cdot \tilde{b}_{\lambda
})+u_{0}\,(s^{\dagger }s^{\dagger }\tilde{s}\tilde{s})-u_{1}\,s^{\dagger
}(b_{\rho }^{\dagger }\cdot \tilde{b}_{\rho }+b_{\lambda }^{\dagger }\cdot 
\tilde{b}_{\lambda })\tilde{s}  \nonumber \\
&&+v_{0}\,\left[ (b_{\rho }^{\dagger }\cdot b_{\rho }^{\dagger }+b_{\lambda
}^{\dagger }\cdot b_{\lambda }^{\dagger })\tilde{s}\tilde{s}+s^{\dagger
}s^{\dagger }(\tilde{b}_{\rho }\cdot \tilde{b}_{\rho }+\tilde{b}_{\lambda
}\cdot \tilde{b}_{\lambda })\right]  \nonumber \\
&&+\sum_{l=0,2}w_{l}\,(b_{\rho }^{\dagger }\times b_{\rho }^{\dagger
}+b_{\lambda }^{\dagger }\times b_{\lambda }^{\dagger })^{(l)}\cdot (\tilde{b%
}_{\rho }\times \tilde{b}_{\rho }+\tilde{b}_{\lambda }\times \tilde{b}%
_{\lambda })^{(l)}  \nonumber \\
&&+\sum_{l=0,2}c_{l}\,\left[ (b_{\rho }^{\dagger }\times b_{\rho }^{\dagger
}-b_{\lambda }^{\dagger }\times b_{\lambda }^{\dagger })^{(l)}\cdot (\tilde{b%
}_{\rho }\times \tilde{b}_{\rho }-\tilde{b}_{\lambda }\times \tilde{b}%
_{\lambda })^{(l)}+4\,(b_{\rho }^{\dagger }\times b_{\lambda }^{\dagger
})^{(l)}\cdot (\tilde{b}_{\lambda }\times \tilde{b}_{\rho })^{(l)}\right] 
\nonumber \\
&&+c_{1}\,(b_{\rho }^{\dagger }\times b_{\lambda }^{\dagger })^{(1)}\cdot (%
\tilde{b}_{\lambda }\times \tilde{b}_{\rho })^{(1)}~,  \label{hs3}
\end{eqnarray}
with $\tilde{b}_{\rho ,m}=(-1)^{1-m}b_{\rho ,-m}$, $\tilde{b}_{\lambda
,m}=(-1)^{1-m}b_{\lambda ,-m}$ and $\tilde{s}=s$. Here the dots indicate
scalar products and the crosses tensor products with respect to the rotation
group. By construction, the Hamiltonian commutes with the number operator 
\ba
\hat{N} &=& s^{\dagger} \tilde{s} 
- b_{\rho}^{\dagger} \cdot \tilde{b}_{\rho} 
- b_{\lambda}^{\dagger} \cdot \tilde{b}_{\lambda} 
\;=\; s^{\dagger}s + \sum_{m} \left( b_{\rho,m}^{\dagger} b_{\rho,m} 
+ b_{\lambda,m}^{\dagger} b_{\lambda,m} \right) ~,
\label{number}
\ea
and therefore conserves the total number of bosons $N=n_{s}+n_{\rho
}+n_{\lambda }$. In addition to $N$, the eigenfunctions have good angular
momentum $L$, parity $P$, and permutation symmetry $t$. The three symmetry
classes of the $S_{3}$ permutation group are characterized by the
irreducible representations: $t=S$ for the one-dimensional symmetric
representation, $t=A$ for the one-dimensional antisymmetric representation,
and $t=M$ for the two-dimensional mixed symmetry representation.

\subsection{Dynamic symmetries}

The $S_{3}$ invariant $U(7)$ mass operator of Eq.~(\ref{hs3}) has a rich
algebraic structure. It is of general interest to study limiting situations,
in which the energy spectra can be obtained in closed form. These special
cases correspond to dynamic symmetries of the Hamiltonian. Dynamic
symmetries arise, when (i) the Hamiltonian has an algebraic structure $G$,
and (ii) it can be expressed in terms of Casimir invariants of a chain of
subalgebras of $G$ only. Here we consider two different dynamic symmetries,
corresponding to the six-dimensional spherical and deformed oscillator cases
discussed in the Schr\"{o}dinger picture in Sect.~3.

For $v_{0}=0$ in Eq.~(\ref{hs3}), we recover the six-dimensional oscillator 
model, since there is no coupling between different harmonic oscillator 
shells. The oscillator is harmonic if all terms except $\epsilon_0$ and 
$\epsilon_1$ are set to zero, otherwise it is anharmonic. This dynamic 
symmetry corresponds to the reduction 
\ba
U(7) \supset U(6) ,\supset \cdots 
\ea
The dots here indicate further algebras needed to classify the states
uniquely. Since our purpose 
here is to present a global analysis of the situation, we do not dwell on the
group theoretical aspects of this problem and rather note that the one-body
Hamiltonian 
\begin{eqnarray}
H_1 &=& -\epsilon_{1} \, (b_{\rho}^{\dagger} \cdot \tilde{b}_{\rho} 
+ b_{\lambda}^{\dagger} \cdot \tilde{b}_{\lambda})  
\nonumber\\
&=& \epsilon_{1} \sum_{m} ( b_{\rho,m}^{\dagger} b_{\rho,m} 
+ b_{\lambda,m}^{\dagger} b_{\lambda,m} )  
\nonumber\\
&=& \epsilon_{1} \, \hat{C}_{1U(6)} ~,  \label{ham1}
\end{eqnarray}
gives the energy spectrum of a six-dimensional spherical oscillator 
\begin{eqnarray}
E_1(n) &=& \epsilon _{1} \, n ~. 
\label{e1}
\end{eqnarray}
Here $\hat{C}_{1U(6)}$ denotes the linear Casimir operator of $U(6)$. 
The label $n$ represents the number of oscillator quanta 
$n=n_{\rho }+n_{\lambda }=0,1,\ldots ,N$. 
This special case is called the $U(6)$ limit.  
In Fig.~\ref{hosc} we show the structure of the spectrum of the spherical
harmonic oscillator with $U(6)$ symmetry. For three identical clusters, the
wave functions must transform as the symmetric representation $t=S$ of the
permutation group $S_{3}$. The levels are grouped into oscillator shells
characterized by $n$. In Fig.~\ref{hosc}, the levels belonging to an
oscillator shell $n$ are further classified by $\sigma=n,n-2,\ldots,1$ 
or $0$ for $n$ odd or even. The quantum number $\sigma$ labels the 
representations of $SO(6)$, a subgroup of $U(6)$. The ground state has 
$n=0$ and $L^{P}=0^{+}$. We
note, that the $n=\sigma =1$ shell is absent, since it does not contain a
symmetric state with $t=S$. The two-phonon multiplet $n=2$ consists of the
states $L^{P}=2^{+}$ with $\sigma =2$ and $0^{+}$ with $\sigma =0$. 
The degeneracy of the harmonic oscillator shells can be split
by adding invariants of subgroups of $U(6)$ \cite{BL2}. 

For the six-dimensional spherical oscillator, the number of oscillator
quanta $n$ is a good quantum number. However, when $v_{0}\neq 0$ in Eq.~(\ref
{hs3}), the oscillator shells with $\Delta n=\pm 2$ are mixed, and the
eigenfunctions are spread over many different oscillator shells. A dynamic
symmetry that involves the mixing between oscillator shells, is provided by
the reduction 
\ba
U(7)\supset SO(7) \supset SO(6) \supset \cdots 
\ea
Again the dots indicate further subalgebras needed to classify the states
uniquely. They can be taken as discussed previously. We consider now an
algebraic Hamiltonian of the form 
\begin{eqnarray}
H_2 &=&\kappa \left[ \hat{N}(\hat{N}+5)-(\hat{D}_{\rho }\cdot \hat{D}%
_{\rho }+\hat{D}_{\lambda }\cdot \hat{D}_{\lambda })\right]  \nonumber \\
&=&\kappa \left[ \hat{N}(\hat{N}+5)-\hat{C}_{2SO(7)}+\hat{C}_{2SO(6)}\right]
\nonumber \\
&=&\kappa \left[ ( s^{\dagger }s^{\dagger }-b_{\rho }^{\dagger }\cdot
b_{\rho }^{\dagger }-b_{\lambda }^{\dagger }\cdot b_{\lambda }^{\dagger
}) \, ( \tilde{s}\tilde{s}-\tilde{b}_{\rho }\cdot \tilde{b}_{\rho
}-\tilde{b}_{\lambda }\cdot \tilde{b}_{\lambda }) +\hat{C}_{2SO(6)}%
\right] ~,  \label{ham2}
\end{eqnarray}
where $\hat N$ is the number operator of Eq.~(\ref{number}) and 
$\hat D$ denote the dipole operators  
\begin{eqnarray}
\hat{D}_{\rho ,m} &=&(b_{\rho }^{\dagger }\times \tilde{s}-s^{\dagger
}\times \tilde{b}_{\rho })_{m}^{(1)}~,  \nonumber \\
\hat{D}_{\lambda ,m} &=&(b_{\lambda }^{\dagger }\times \tilde{s}-s^{\dagger
}\times \tilde{b}_{\lambda })_{m}^{(1)}~.
\label{dipole}
\end{eqnarray}
The operators $\hat{C}_{2SO(7)}$ and $\hat{C}_{2SO(6)}$ represent the 
quadratic Casimir operators of $SO(7)$ and $SO(6)$, respectively. 
The energy spectrum in this case, called the $SO(7)$ limit, is given by 
the eigenvalues of the Casimir operators 
$\langle \hat{C}_{2SO(7)}\rangle =\omega(\omega +5)$ and 
$\langle \hat{C}_{2SO(6)}\rangle =\sigma(\sigma +4)$  
\ba
E_2(\omega,\sigma) &=& \kappa \, [(N-\omega)(N+\omega +5) 
+\sigma (\sigma+4) ] ~. 
\label{e2}
\ea
Here $\omega =N,N-2,\ldots,1$ or $0$ for $N$ odd or even, respectively, 
labels the symmetric representations of $SO(7)$, and 
$\sigma=0,2,3,\ldots,\omega$ those of $SO(6)$ (note that $\sigma=1$ 
is missing, since it doesnot contain a symmetric state). 
The correspondence with the energy spectrum
of the deformed oscillator of Eq.~(\ref{ener2}) can be seen by introducing a
vibrational quantum number $v=(N-\omega )/2$. The energy formula reduces to 
\begin{eqnarray}
E_2(v,\sigma) &=& 4\kappa N \, v \left( 1-\frac{2v-5}{2N} \right) 
+ \kappa \, \sigma(\sigma +4) ~,
\label{ener}
\end{eqnarray}
which to leading order in $N$, is linear in the vibrational quantum number.
In Fig.~\ref{defosc} we show the spectrum of the deformed oscillator with $%
SO(7)$ symmetry. The states are now ordered in bands labeled by $\omega $,
rather than in harmonic oscillator shells. Although the size of the model
space, and hence the total number of states, is the same as for the harmonic
oscillator, the ordering and classification of the states is different. For
example, in the $U(6)$ limit all states are vibrational, whereas the $SO(7)$
limit gives rise to a rotational-vibrational spectrum, characterized by a
series of rotational bands which are labeled by $\omega $, or equivalently
by the vibrational quantum number $v=(N-\omega )/2=0,1,\ldots $~.

\subsection{Oblate symmetric top}

This situation does not correspond to a dynamic symmetry of the Hamiltonian $%
H$. However it is relatively simple to construct an algebraic Hamiltonian
that produces the spectrum of an oblate symmetric top. This Hamiltonian was
suggested in our preliminary account of this work \cite{BI}. It can be
written as
\begin{eqnarray}
H_3 &=& \xi_{1} \, ( s^{\dagger }s^{\dagger }-b_{\rho }^{\dagger
}\cdot b_{\rho }^{\dagger }-b_{\lambda }^{\dagger }\cdot b_{\lambda
}^{\dagger }) ( \tilde{s}\tilde{s}-\tilde{b}_{\rho }\cdot \tilde{b%
}_{\rho }-\tilde{b}_{\lambda }\cdot \tilde{b}_{\lambda })  \nonumber \\
&&+\xi _{2} \left[( b_{\rho }^{\dagger }\cdot b_{\rho }^{\dagger }-b_{\lambda
}^{\dagger }\cdot b_{\lambda }^{\dagger }) ( \tilde{b}_{\rho
}\cdot \tilde{b}_{\rho }-\tilde{b}_{\lambda }\cdot \tilde{b}_{\lambda
})  +4( b_{\rho }^{\dagger }\cdot b_{\lambda }^{\dagger }) 
( \tilde{b}_{\lambda }\cdot \tilde{b}_{\rho }) \right]  \nonumber \\
&&+2\kappa_1 \, (b_{\rho }^{\dagger} \times \tilde{b}_{\rho} 
+ b_{\lambda}^{\dagger} \times \tilde{b}_{\lambda})^{(1)} \cdot 
(b_{\rho}^{\dagger} \times \tilde{b}_{\rho}+b_{\lambda}^{\dagger} \times 
\tilde{b}_{\lambda})^{(1)}  \nonumber \\
&&+3\kappa_2 \, (b_{\rho}^{\dagger} \times \tilde{b}_{\lambda}
-b_{\lambda}^{\dagger} \times \tilde{b}_{\rho}) ^{(0)} \cdot 
(b_{\lambda}^{\dagger} \times \tilde{b}_{\rho}-b_{\rho}^{\dagger} \times 
\tilde{b}_{\lambda})^{(0)} ~.
\label{ham3}
\end{eqnarray}
Since this Hamiltonian does not correspond to a dynamic symmetry, its
spectrum must be obtained by diagonalizing it in an appropriate basis. A
computer program has been written for this purpose by one of us (RB) and is
available upon request \cite{acm}. 
An approximate energy formula can be obtained by making
use of the method of intrinsic or coherent states as discussed in the
following section. 

\section{Geometric analysis}

The geometric properties of the general algebraic Hamiltonian $H$ of 
Eq.~(\ref{hs3}) can be studied by using intrinsic or coherent states 
\cite{cs}. The ground state coherent state for the problem at hand can
be written as 
\begin{equation}
|N;\vec{\alpha}_{\rho },\vec{\alpha}_{\lambda }\rangle =\frac{1}{\sqrt{N!}}%
(b_{c}^{\dagger })^{N}\,|0\rangle ~,
\end{equation}
with 
\begin{equation}
b_{c}^{\dagger }=\sqrt{1-\vec{\alpha}_{\rho }\cdot \vec{\alpha}_{\rho
}^{\,\ast }-\vec{\alpha}_{\lambda }\cdot \vec{\alpha}_{\lambda }^{\,\ast }}%
\;s^{\dagger }+\vec{\alpha}_{\rho }\cdot \vec{b}_{\rho }^{\,\dagger }+\vec{%
\alpha}_{\lambda }\cdot \vec{b}_{\lambda }^{\,\dagger }~.
\label{bc}
\end{equation}
The condensate boson Eq.~(\ref{bc}) is parametrized in terms of two vectors, 
$\vec{\alpha}_{\rho }$ and $\vec{\alpha}_{\lambda }$, which can be
transformed to intrinsic coordinates $q_{\rho }$, $\theta _{\rho }$, $\phi
_{\rho }$, and $q_{\lambda }$, $\theta _{\lambda }$, $\phi _{\lambda }$, and
their conjugate momenta \cite{onno} 
\begin{equation}
\alpha _{k,\mu }=\frac{1}{\sqrt{2}}\sum_{\nu }{\cal D}_{\mu \nu }^{(1)}(\phi
_{k},\theta _{k},0)\,\beta _{k,\nu }~,
\end{equation}
with 
\begin{equation}
\left( 
\begin{array}{c}
\beta _{k,1} \\ 
\beta _{k,0} \\ 
\beta _{k,-1}
\end{array}
\right) =\left( 
\begin{array}{c}
\,[-p_{\phi _{k}}/\sin \theta _{k}-ip_{\theta _{k}}]/q_{k}\sqrt{2} \\ 
q_{k}+ip_{k} \\ 
\,[-p_{\phi _{k}}/\sin \theta _{k}+ip_{\theta _{k}}]/q_{k}\sqrt{2}
\end{array}
\right) ~,
\end{equation}
with $k=\rho $, $\lambda $. The classical limit of a normal ordered operator
can be defined as the coherent state expectation value of the operator
divided by the number of bosons $N$.

\subsection{Dynamic symmetries}

We start by analyzing the two dynamic symmetries discussed
in the previous section.
For the $U(6)$ limit, the classical limit of the Hamiltonian is given by 
\begin{eqnarray}
H_{1,{\rm cl}} &=&\frac{1}{N}\langle N;\vec{\alpha}_{\rho },\vec{\alpha}%
_{\lambda }\mid \,: H_1 :\,\mid N;\vec{\alpha}_{\rho },\vec{\alpha}_{\lambda
}\rangle  \nonumber \\
&=& \epsilon_{1} \, (\vec{\alpha}_{\rho }\cdot \vec{\alpha}%
_{\rho }^{\,\ast }+\vec{\alpha}_{\lambda }\cdot \vec{\alpha}_{\lambda
}^{\,\ast })  \nonumber \\
&=& \epsilon_{1} \frac{1}{2}\left( p_{\rho }^{2}+q_{\rho
}^{2}+\vec{L}_{\rho }^{2}/q_{\rho }^{2}+p_{\lambda }^{2}+q_{\lambda }^{2}+%
\vec{L}_{\lambda }^{2}/q_{\lambda }^{2}\right) ~,  \label{hcl1}
\end{eqnarray}
where $\vec{L}_{k}^{2}$ is the angular momentum in polar coordinates 
\ba
\vec{L}_{k}^{2}&=&p_{\theta _{k}}^{2}+\frac{p_{\phi _{k}}^{2}}{\sin ^{2}\theta
_{k}}~,
\ea
with $k=\rho $, $\lambda $. Eq.~(\ref{hcl1}) confirms the interpretation of
the $U(6)$ limit in terms of a six-dimensional spherical oscillator.

For the $SO(7)$ limit, we find 
\begin{eqnarray}
H_{2,{\rm cl}} 
&=&\kappa (N-1)\left[ q^{2}p^{2}+(1-q^{2})^{2}-\frac{\partial ^{2}}{\partial
\chi ^{2}}-4\cot 2\chi \frac{\partial }{\partial \chi }+\frac{\vec{L}_{\rho
}^{2}}{\sin ^{2}\chi }+\frac{\vec{L}_{\lambda }^{2}}{\cos ^{2}\chi }\right]
~.
\end{eqnarray}
Here we have made a change to hyperspherical variables 
\begin{equation}
q_{\rho }\;=\;q\sin \chi ~,\hspace{1cm}q_{\lambda }\;=\;q\cos \chi ~,
\end{equation}
and their conjugate momenta. In the limit of small oscillations around the
minimum of the potential energy surface $q=1+\Delta q$ we find to leading
order in $N$ a one-dimensional harmonic oscillator in the hyperradius $q$ 
\ba
H_{2,{\rm cl}} &\rightarrow& \kappa N \left( p^{2}+4(\Delta q)^{2}\right) ~,
\ea
with frequency $4\kappa N$. To leading order in $N$, the frequency
coincides with that of Eq.~(\ref{ener}). This analysis shown the connection
between the $SO(7)$ dynamical symmetry and the displaced oscillator.

\subsection{Oblate symmetric top}

For the oblate symmetric top, the method of intrinsic states becomes 
particularly useful, since this case does not correpond to a dynamic 
symmetry. We first consider the vibrational part of the Hamiltonian of 
Eq.~(\ref{ham3}) 
\begin{eqnarray}
H_{3,{\rm vib}} &=&\xi _{1}\,(R^{2}\,s^{\dagger }s^{\dagger }-b_{\rho
}^{\dagger }\cdot b_{\rho }^{\dagger }-b_{\lambda }^{\dagger }\cdot
b_{\lambda }^{\dagger })\,(R^{2}\,\tilde{s}\tilde{s}-\tilde{b}_{\rho }\cdot 
\tilde{b}_{\rho }-\tilde{b}_{\lambda }\cdot \tilde{b}_{\lambda })  
\nonumber\\
&&+\xi _{2}\,\left[ (b_{\rho }^{\dagger }\cdot b_{\rho }^{\dagger
}-b_{\lambda }^{\dagger }\cdot b_{\lambda }^{\dagger })\,(\tilde{b}_{\rho
}\cdot \tilde{b}_{\rho }-\tilde{b}_{\lambda }\cdot \tilde{b}_{\lambda
})+4\,(b_{\rho }^{\dagger }\cdot b_{\lambda }^{\dagger })\,(\tilde{b}%
_{\lambda }\cdot \tilde{b}_{\rho })\right] ~.  \label{oblate}
\end{eqnarray}
For $R^{2}=0$, this Hamiltonian has $U(7)\supset U(6)$ symmetry and
corresponds to a spherical vibrator, whereas for $R^{2}=1$ and $\xi _{2}=0$
it has $U(7)\supset SO(7)$ symmetry and corresponds to a deformed
oscillator. Here we analyze the general case with $R^{2}\neq 0$ and $\xi
_{1} $, $\xi _{2}>0$. The classical limit of the Hamiltonian of Eq.~(\ref
{oblate}) has a complicated structure. The equilibrium configuration of the
potential energy surface is given by 
\begin{equation}
q_{0}\;=\;\sqrt{2R^{2}/(1+R^{2})}~,\hspace{1cm}\chi _{0}\;=\;\pi /4~,\hspace{%
1cm}\zeta _{0}\;=\;\pi /4~,
\end{equation}
where the relative angle between $\vec{\alpha}_{\rho }$ and $\vec{\alpha}%
_{\lambda }$ is denoted by $2\zeta $. In the limit of small oscillations
around the minimum $q=q_{0}+\Delta q$, $\chi =\chi _{0}+\Delta \chi $ and $%
\zeta =\zeta _{0}+\Delta \zeta $, the intrinsic degrees of freedom $q$, $%
\chi $ and $\zeta $ decouple and become harmonic. To leading order in $N$ we
find 
\begin{eqnarray}
H_{3,{\rm cl}} &=&\xi _{1}N\left[ \frac{2R^{2}}{1+R^{2}}p^{2}+2R^{2}(1+R^{2})(%
\Delta q)^{2}\right]  \nonumber \\
&&+\xi _{2}N\left[ p_{\chi }^{2}+\frac{4R^{4}}{(1+R^{2})^{2}}(\Delta \chi
)^{2}+p_{\zeta }^{2}+\frac{4R^{4}}{(1+R^{2})^{2}}(\Delta \zeta )^{2}\right]
~.
\end{eqnarray}
Standard quantization of the harmonic oscillator yields the vibrational
energy spectrum of an oblate top 
\ba
E_{3,{\rm vib}}(v_{1},v_{2})&=&\omega_{1}(v_{1}+\frac{1}{2}) 
+ \omega_{2}(v_{2}+1) ~,
\ea
with frequencies 
\begin{equation}
\omega _{1} \;=\; 4NR^{2} \xi_{1} ~,  \hspace{1cm} 
\omega _{2} \;=\; \frac{4NR^{2}}{1+R^{2}} \xi_{2} ~,
\end{equation}
in agreement with the results obtained in a normal mode analysis \cite{BIL}. 
Here $v_{1}$ and $v_{2}$ represent vibrational quantum numbers for a one- 
and two-dimensional harmonic oscillator, respectively. 

We consider next the rotational part of the Hamiltonian which can be
rewritten as 
\ba
H_{3,{\rm rot}} &=& 2\kappa_1 \, (b_{\rho }^{\dagger} \times \tilde{b}_{\rho} 
+ b_{\lambda}^{\dagger} \times \tilde{b}_{\lambda})^{(1)} \cdot 
(b_{\rho}^{\dagger} \times \tilde{b}_{\rho}+b_{\lambda}^{\dagger} \times 
\tilde{b}_{\lambda})^{(1)}  \nonumber\\
&&+3\kappa_2 \, (b_{\rho}^{\dagger} \times \tilde{b}_{\lambda}
-b_{\lambda}^{\dagger} \times \tilde{b}_{\rho}) ^{(0)} \cdot 
(b_{\lambda}^{\dagger} \times \tilde{b}_{\rho}-b_{\rho}^{\dagger} \times 
\tilde{b}_{\lambda})^{(0)} ~. 
\ea
Both terms commute with the general $S_{3}$ invariant Hamiltonian of Eq.~(%
\ref{hs3}) and hence correspond to exact symmetries. The eigenvalues are
given by 
\begin{eqnarray}
E_{3,{\rm rot}} &=& \kappa_{1} \, L(L+1) + \kappa_{2} \, m_{F}^{2} 
\nonumber\\
&=& \kappa_{1} \, L(L+1) + \kappa_{2} \, (K \mp 2\ell_{2})^{2} ~.
\end{eqnarray}
Here we have used, that for the oblate top the quantum number $m_{F}$ is 
related to the projection $K$ of the angular momentum on the symmetry-axis 
and $\ell_{2}$ \cite{BDL}. 
The last term contains the effects of the Coriolis force which gives rise to
a $8\kappa_{2} K \ell_{2}$ splitting of the rotational levels.

\begin{figure}
\centering
\setlength{\unitlength}{1.0pt} 
\begin{picture}(310,300)(0,0)
\thinlines
\put (  0,  0) {\line(1,0){310}}
\put (  0,300) {\line(1,0){310}}
\put (  0,  0) {\line(0,1){300}}
\put (310,  0) {\line(0,1){300}}
\thicklines
\put ( 30, 60) {\line(1,0){20}}
\put ( 30, 96) {\line(1,0){20}}
\put ( 30,180) {\line(1,0){20}}
\put ( 70,132) {\line(1,0){20}}
\put ( 70,180) {\line(1,0){20}}
\multiput ( 40,200)(0,5){5}{\circle*{0.1}}
\multiput ( 80,200)(0,5){5}{\circle*{0.1}}
\multiput ( 70, 60)(5,0){42}{\circle*{0.1}}
\multiput ( 80, 60)(0,5){15}{\circle*{0.1}}
\thinlines
%\put ( 20, 40) {$K$}
\put ( 40, 40) {$0$}
\put ( 80, 40) {$3$}
%\put ( 20, 25) {$(v_1,v_2^{\ell_2})$}
\put ( 50, 25) {$(0,0^0)$}
\put ( 55, 55) {$0^+$}
\put ( 55, 91) {$2^+$}
\put ( 55,175) {$4^+$}
\put ( 95,127) {$3^-$}
\put ( 95,175) {$4^-$}
\thicklines
\put (120, 90) {\line(1,0){20}}
\put (120,126) {\line(1,0){20}}
\put (120,210) {\line(1,0){20}}
\put (160,162) {\line(1,0){20}}
\put (160,210) {\line(1,0){20}}
\multiput (130,230)(0,5){5}{\circle*{0.1}}
\multiput (170,230)(0,5){5}{\circle*{0.1}}
\multiput (130, 60)(0,5){7}{\circle*{0.1}}
\multiput (170, 60)(0,5){21}{\circle*{0.1}}
\thinlines
\put (130, 40) {$0$}
\put (170, 40) {$3$}
\put (140, 25) {$(1,0^0)$}
\put (145, 85) {$0^+$}
\put (145,121) {$2^+$}
\put (145,205) {$4^+$}
\put (185,157) {$3^-$}
\put (185,205) {$4^-$}
\thicklines
\put (210,170) {\line(1,0){20}}
\put (210,194) {\line(1,0){20}}
\put (210,230) {\line(1,0){20}}
\put (250,194) {\line(1,0){20}}
\put (250,230) {\line(1,0){20}}
\multiput (220,250)(0,5){5}{\circle*{0.1}}
\multiput (260,250)(0,5){5}{\circle*{0.1}}
\multiput (220, 60)(0,5){23}{\circle*{0.1}}
\multiput (260, 60)(0,5){27}{\circle*{0.1}}
\thinlines
\put (220, 40) {$1$}
\put (260, 40) {$2$}
\put (230, 25) {$(0,1^1)$}
\put (235,165) {$1^-$}
\put (235,189) {$2^-$}
\put (235,225) {$3^-$}
\put (275,189) {$2^+$}
\put (275,225) {$3^+$}
\end{picture}
\vspace{1cm}
\caption[Spectrum: oblate top]
{Schematic spectrum of an oblate symmetric top. All
states are symmetric under $S_3$.}
\label{top}
\end{figure}
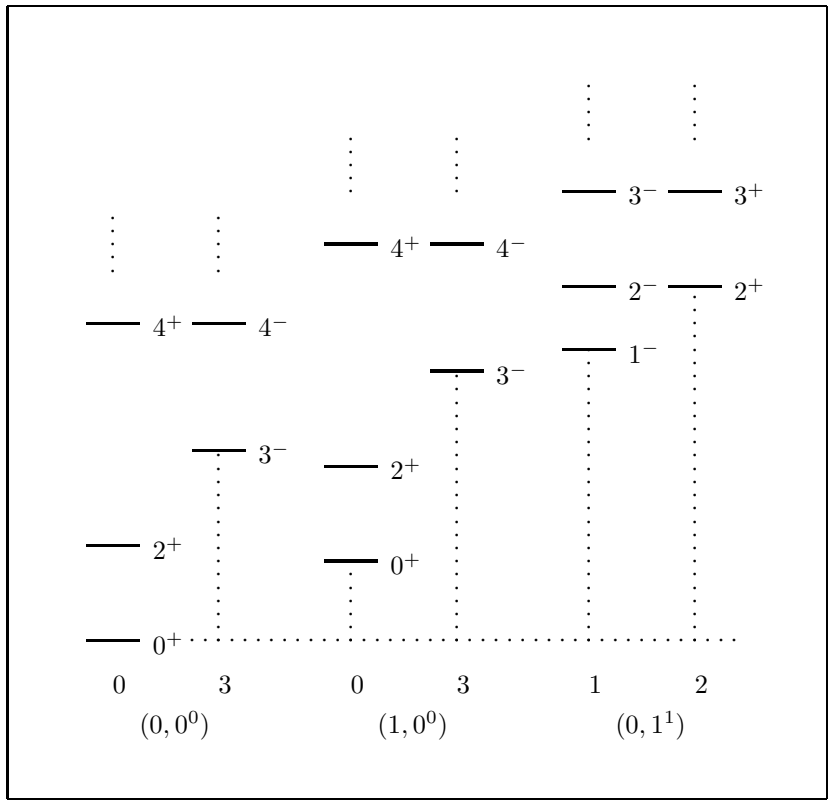

The total (rotation-vibration) wave functions of a rigid triangular
configuration can be written as 
\begin{equation}
\mid (v_{1},v_{2}^{\ell _{2}});K,L_{t}^{P},M\rangle .
\end{equation}
For three identical particles, the wave functions for rigid configurations
must transform as the symmetric representations $A_{1}$ of ${\cal D}_{3}\sim
S_{3}$. This imposes some conditions on the allowed values of the angular
momenta. For vibrational bands with $\ell _{2}=0$ and $1$, the allowed
values of the angular momenta are 
\ba
\begin{array}{lll}
(v_{1},v_{2}^{\ell _{2}=0})\;:\; & K=3n & n=0,1,2,\ldots \\ 
& L=0,2,4,\ldots & \mbox{ for }K=0 \\ 
& L=K,K+1,K+2,\ldots & \mbox{ for }K\neq 0 \\ 
(v_{1},v_{2}^{\ell _{2}=1})\;:\; & K=3n+1,3n+2 & n=0,1,2,\ldots \\ 
& L=K,K+1,K+2,\ldots & 
\end{array}
\ea
The parity is $P=(-)^{K}$. The vibrational band $(1,0^{0})$ has the same
angular momenta $L^{P}=0^{+},2^{+},3^{-},4^{\pm },\ldots$~, as the ground
state band $(0,0^{0})$, while the angular momentum content of the doubly
degenerate vibration $(0,1^{1})$ is given by $L^{P}=1^{-},2^{\mp },3^{\mp
},\ldots$~. Combining all results one can write an approximate expression
for the energy eigenvalues fo the oblate symmetric top
\ba
E(v_{1},v_{2}^{\ell_{2}},K,L,M) &=& E_0 + \omega_{1}(v_{1}+\frac{1}{2}) 
+ \omega_{2}(v_{2}+1) 
\nonumber\\
&&+ \kappa_{1} \, L(L+1) + \kappa_{2} \, (K \mp 2\ell_{2})^{2} ~.
\label{ost}
\ea
In Fig.~\ref{top} we show the structure of the spectrum of the oblate top 
according to the approximate energy formula of Eq.~(\ref{ost}). 
The energy spectrum consists of a series of rotational bands labeled by 
$(v_1,v_2^{\ell_2})$ and $K$. The degeneracy between states with 
different values of $K$ can be split by the last term in Eq.~(\ref{ost}). 

\section{Transition probabilities}

In order to calculate transition form factors in the algebraic cluster 
model one has to express the transition operator in terms of the algebraic 
operators. In the large $N$ limit, the $U(7)$ dipole operators 
$\hat D_{\rho}$ and $\hat D_{\lambda}$ of Eq.~(\ref{dipole}) correspond to 
the Jacobi coordinates \cite{onno}. The matrix elements of 
$\exp(-iq \sqrt{2/3} \lambda_{z})$ can be obtained algebraically 
by making the replacement \cite{BIL} 
\ba
\sqrt{2/3} \lambda_z &\rightarrow& \beta \hat D_{\lambda,z}/X_D ~, 
\ea
where $\beta$ represents the scale of the coordinate and $X_D$ is given 
by the reduced matrix element of the dipole operator \cite{BIL}. 
In summary, the transition form factors can be expressed in the ACM as 
\ba
{\cal F}(i \rightarrow f;q) &=& \langle \gamma_f,L_f,M \, 
| \, \hat T \, | \, \gamma_i,L_i,M \rangle ~, 
\ea
with
\ba
\hat T &=& e^{i \epsilon \hat{D}_{\lambda,z}}
\;=\; e^{-iq\beta \hat{D}_{\lambda,z}/X_{D}} ~. 
\label{trans}
\ea
For the general ACM Hamiltonian of Eq.~(\ref{hs3}), 
the form factors cannot be obtained in closed analytic form, but have to 
be calculated numerically. Hereto, a computer program has been developed 
\cite{acm}, in which the form factors are obtained exactly by using 
the symmetry properties of the transition operator of Eq.~(\ref{trans}) 
(see Appendix D of \cite{BIL}). 

\subsection{Dynamic symmetries}

When a dynamic symmetry occurs the matrix elements of the transition 
operators can be obtained in explicit analytic form. The general procedure 
to derive the transition form factors in the $U(6)$ and $SO(7)$ limits 
has been discussed in \cite{BG,BIL}. 

\begin{figure}
\centerline{\hbox{\epsfig{figure=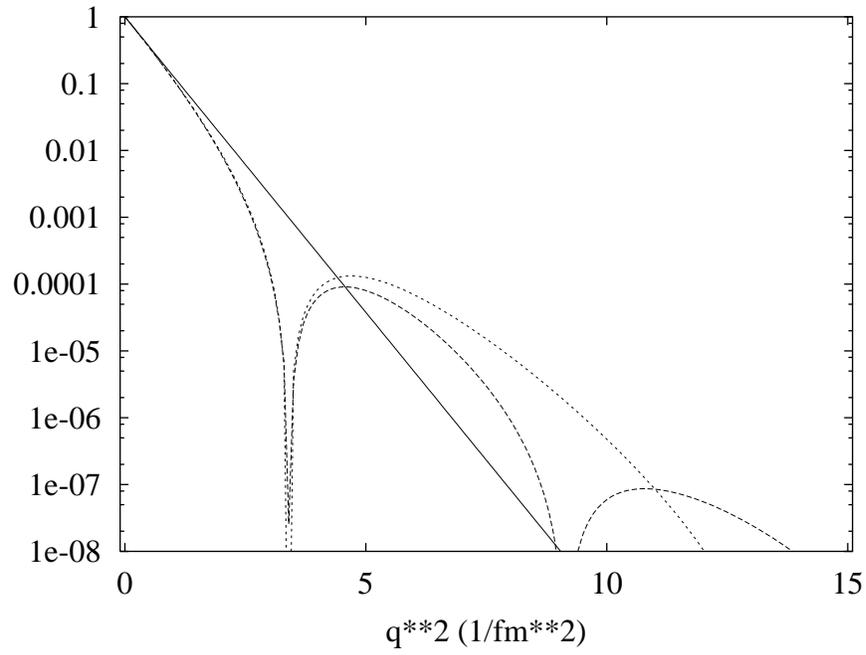} }}
\caption[Elastic form factor]{Comparison of the elastic form factor 
$|{\cal F}(0^+_1 \rightarrow 0^+_1;q)|^2$, calculated for 
$N \rightarrow \infty$ 
(a) in $U(6)$ with $\beta=1.70$ fm and $\alpha=0.47$ fm$^{-2}$ (solid line), 
(b) in $SO(7)$ with $\beta=1.97$ fm and $\alpha=0.67$ fm$^{-2}$ (dashed line), 
(c) in the oblate top with $\beta=1.70$ fm and $\alpha=0.47$ fm$^{-2}$ 
(dotted line).}
\label{ffel}
\end{figure}

\begin{figure}
\centerline{\hbox{\epsfig{figure=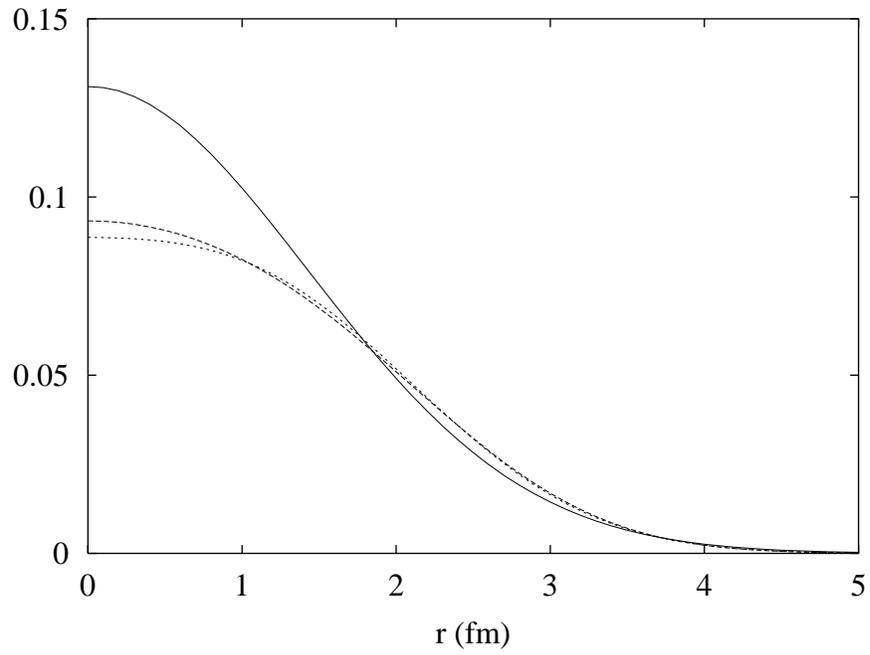} }}
\caption[Charge distribution]
{Ground state charge distribution calculated for $N \rightarrow \infty$ 
(a) in $U(6)$ with $\beta=1.70$ fm and $\alpha=0.47$ fm$^{-2}$ (solid line), 
(b) in $SO(7)$ with $\beta=1.97$ fm and $\alpha=0.67$ fm$^{-2}$ (dashed line), 
(c) in the oblate top with $\beta=1.70$ fm and $\alpha=0.47$ fm$^{-2}$ 
(dotted line).}
\label{chd}
\end{figure}

In the $U(6)$ limit, the elastic form factor is given by \cite{BIL} 
\ba
{\cal F}(0^+_1 \rightarrow 0^+_1;q) &=& (\cos \epsilon)^N   
\; \rightarrow \; e^{-q^2 \beta^2/6} ~.
\label{ffel1}
\ea
In the last step we have taken the large $N$ limit, such that 
$\epsilon X_D=\epsilon \sqrt{3N}=-q\beta$ remains finite. 
For the extended charge distribution, the form factor of Eq.~(\ref{ffel1}) 
is multiplied by a Gaussian $\exp(-q^2/4\alpha)$. The coefficients $\beta$ 
and $\alpha$ cannot be determined independently, since both have the same 
dependence on $q^2$. The charge radius can be used to fix one particular 
combination
\ba
\langle r^{2} \rangle^{1/2} &=& \sqrt{\frac{3}{2\alpha}+\beta^{2}} ~. 
\label{radius}
\ea
In Table~\ref{ffu6} we show the results for some transition form factors 
in the $U(6)$ limit of the ACM. In the large $N$ limit, 
they reduce to those obtained for the harmonic oscillator of Table~\ref{ff}. 
All form factors exhibit an exponential fall-off with momentum transfer. 
In Fig.~\ref{ffel} we show the elastic form factor in the large $N$ 
limit (solid line). The coefficients $\beta$ and $\alpha$ were 
taken to give the charge radius of $^{12}$C: 
$\langle r^2 \rangle^{1/2}=2.468$ fm \cite{reuter}. The corresponding 
charge distribution shows an exponential decay with $r$ (solid line in 
Fig.~\ref{chd}).  

For the $SO(7)$ symmetry, the transition form factors can be expressed in 
terms of Gegenbauer polynomials or, equivalently, hypergeometric funcions. 
For the elastic form factor we have \cite{BL2} 
\ba
{\cal F}(0^+_1 \rightarrow 0^+_1;q) &=& \frac{4!N!}{(N+4)!} \, 
C^{(5/2)}_N(\cos \epsilon) 
\nonumber\\
&=& _2F_1(-\frac{N}{2},\frac{N+5}{2},3;\sin^2 \epsilon) 
\; \rightarrow \; \frac{4 J_2(q\beta \sqrt{2})}{q^2 \beta^2} ~. 
\ea
The normalization constant is given by $X_D=\sqrt{N(N+5)/2}$. 
Again, in the last step we have taken the large $N$ limit, such that  
$\epsilon X_D=-q\beta$ remains finite. The $SO(7)$ transition form factors 
are presented in Table~\ref{ffso7}. In the large $N$ limit, they reduce to 
those obtained for the deformed oscillator of Table~\ref{ff}.    
All form factors exhibit an oscillatory behavior. 
The transition form factor to the first excited $0^+$ state which belongs 
to a vibrational excitation $v=(N-\omega)/2=1$ vanishes identically, since 
the dipole operator is a generator of $SO(7)$. 
In Fig.~\ref{ffel} we show the 
elastic form factor for $N \rightarrow \infty$ (dashed line). The values of 
$\beta$ and $\alpha$ are taken to reproduce the minimum in the elastic 
form factor of $^{12}$C at $q^2=3.4$ fm$^{-2}$, and the $^{12}$C charge 
radius of $2.468$ fm. The corresponding charge distribution for the 
deformed oscillator is given by the dashed line in Fig.~\ref{chd}. 

\subsection{Oblate symmetric top}

For the oblate symmetric top, the form factors 
can only be obtained in closed form in the large $N$ limit \cite{BIL} 
(see Table~\ref{ff}). As an example, the elastic form factor has been 
obtained as \cite{BIL,Inopin}
\ba
{\cal F}(0^+_1 \rightarrow 0^+_1;q) &\rightarrow& j_0(q \beta) ~. 
\ea
In this case, the normalization constant is given by 
$X_D=NR\sqrt{2}/(1+R^2)$. As can be seen from the last column of 
Table~\ref{ff}, all form factors show an oscillatory behavior. 
In Fig.~\ref{ffel} we show the elastic 
form factor in the large $N$ limit (dotted line). 
The coefficients $\beta$ and $\alpha$ have been determined to give the 
same value of the minimum and the charge radius as for the deformed 
oscillator. The charge distribution for the oblate top is given by 
Eq.~(\ref{otch}), and is represented by the dotted line in Fig.~\ref{chd}. 

\begin{figure}
\centerline{\hbox{\epsfig{figure=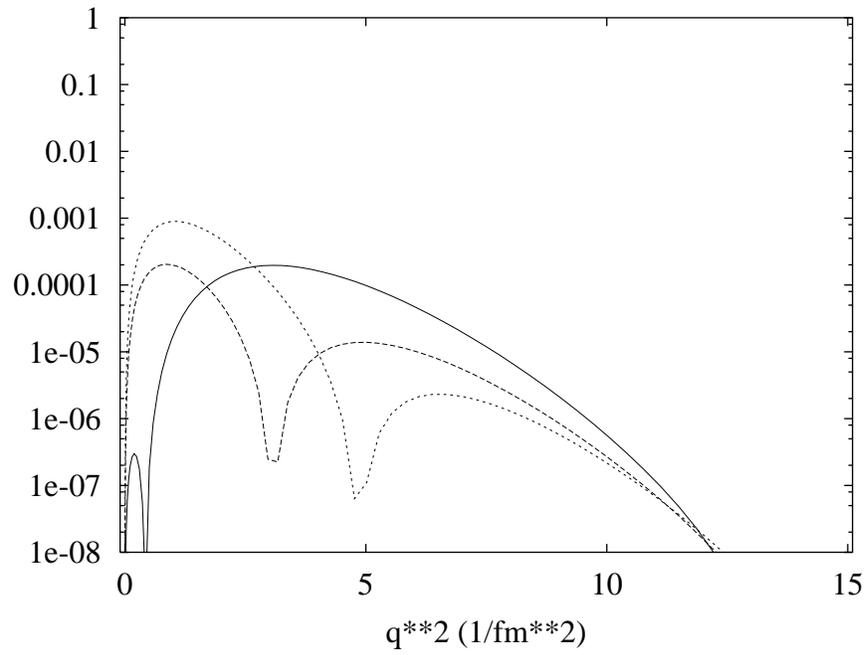} }}
\caption[Inelastic form factor]{Inelastic form factor 
$|{\cal F}(0^+_1 \rightarrow 0^+_2;q)|^2$ for the oblate top, calculated 
for $N=10$ and $R^2=0.5$ (solid line), $1.0$ (dashed line) and 
$1.5$ (dotted line).}
\label{ffinel}
\end{figure}

The transition form factors for vibrational excitations are proportional 
to the corresponding intrinsic transition matrix elements $\chi_1$ and 
$\chi_2$ for each type of vibration $v_1$ and $v_2$, respectively \cite{BI}
\ba 
\chi_1 &=& \frac{1-R^{2}}{2R\sqrt{N}} ~, 
\nonumber\\
\chi_2 &=& \frac{\sqrt{1+R^{2}}}{R\sqrt{2N}}~. 
\label{chi}
\ea
In Fig.~\ref{ffinel} we show the inelastic form factor 
$|{\cal F}(0^+_1 \rightarrow 0^+_2;q)|^2$ 
for different values of $R^2$. As before, the coefficients $\beta$ and 
$\alpha$ are determined by the first minimum of the elastic form 
factor and the charge radius. Whereas the elastic form factor does not 
depend on $R^2$, the inelastic form factor is very sensitive to the value 
of $R^2$, especially with respect to the position of its minimum.  

\section{The nucleus $^{12}$C}

The structure of $^{12}$C has been extensively investigated since the early
days of Nuclear Physics. Comprehensive calculations of $^{12}$C have been
done within the framework of the shell model, starting with the early
calculation of Cohen and Kurath \cite{CK}  
in the $p$-shell and ending with the more 
recent calculations with large shell mixing. $^{12}$C has also been the
battleground for investigations within the framework of the microscopic
cluster model. Here we investigate the extent to which properties of the
low-lying spectrum of $^{12}$C can be described in terms of the macroscopic
cluster model. The reason why a detailed analysis is now feasible is
two-fold. First,we have a set of explicit analytic formulas for energies,
electromagnetic transition rates and form factors for the oblate top that
can be easily be compared with experiments. Second and foremost is that the
development of the algebraic cluster model (ACM) allows us to study in a
straighforward way complex situations, such as those that are not rigid, but
correspond for example to situations which are intermediate between two
limiting cases.

The differences between the three special solutions of the ACM, 
the $U(6)$ limit, the $SO(7)$ limit and the oblate top, are most 
pronounced in the transition form factors. The elastic form factor 
of $^{12}$C shows a minimum at $q^2=3.4$ fm$^{-2}$ and no further 
structure up to about $q^2 \sim 14$ fm$^{-2}$. Also the inelastic form 
factor for the transition $0^+_1 \rightarrow 0^+_2$ shows a clear 
minimum at $\sim q^2=4.5$ fm$^{-2}$. The analysis of the form factors 
that was presented in the previous section shows clearly that 
only the oblate top scenario can account for the features of the 
empirical form factors of $^{12}$C. In the $U(6)$ limit the form 
factors drop off exponentially and have no minima, whereas in the 
$SO(7)$  limit the inelastic form factor vanishes identically. 

Therefore, in this section we analyze the spectroscopy of $^{12}$C 
in the oblate top limit of the ACM. 

\subsection{Energies}

\begin{figure}
\centering
\vspace{15pt}
\setlength{\unitlength}{0.8pt}
\begin{picture}(520,250)(-60,30)
\thinlines
\put (  0, 35) {\line(0,1){215}}
\put (230, 35) {\line(0,1){215}}
\put (460, 35) {\line(0,1){215}}
\put (  0, 35) {\line(1,0){460}}
\put (  0,250) {\line(1,0){460}}
\thicklines
\put (  0, 60) {\line(1,0){5}}
\put (  0,110) {\line(1,0){5}}
\put (  0,160) {\line(1,0){5}}
\put (  0,210) {\line(1,0){5}}
\put (225, 60) {\line(1,0){10}}
\put (225,110) {\line(1,0){10}}
\put (225,160) {\line(1,0){10}}
\put (225,210) {\line(1,0){10}}
\put (455, 60) {\line(1,0){5}}
\put (455,110) {\line(1,0){5}}
\put (455,160) {\line(1,0){5}}
\put (455,210) {\line(1,0){5}}
\put (-20, 55) { 0}
\put (-20,105) { 5}
\put (-20,155) {10}
\put (-20,205) {15}
\put (-60,180) {E(MeV)}
\put (170, 55) {EXP}
\put (400, 55) {TH}
\put ( 20, 60.0) {\line(1,0){20}}
\put ( 20,104.4) {\line(1,0){20}}
\put ( 20,156.4) {\line(1,0){20}}
\put ( 20,200.8) {\line(1,0){20}}
\thinlines
\put ( 45, 55.0) {$0^+$}
\put ( 45, 99.4) {$2^+$}
\put ( 45,151.4) {$3^-$}
\put ( 45,195.8) {$4^+$}
\thicklines
\put ( 70,136.5) {\line(1,0){20}}
\put ( 70,171.6) {\line(1,0){20}}
\thinlines
\put ( 95,131.5) {$0^+$}
\put ( 95,168.6) {$(2^+)$}
\thicklines
\put (120,163.0) {\line(1,0){20}}
\thinlines
\put (145,158.0) {$(0^+)$}
\thicklines
\put (170,168.4) {\line(1,0){20}}
\put (170,178.3) {\line(1,0){20}}
\thinlines
\put (195,163.4) {$1^-$}
\put (195,178.3) {$2^-$}
\thicklines
\put (250, 60.0) {\line(1,0){20}}
\put (250,102.4) {\line(1,0){20}}
\put (250,156.3) {\line(1,0){20}}
\put (250,201.4) {\line(1,0){20}}
\thinlines
\put (275, 55.0) {$0^+$}
\put (275, 97.4) {$2^+$}
\put (275,151.3) {$3^-$}
\put (275,196.4) {$4^+$}
\thicklines
\put (300,136.5) {\line(1,0){20}}
\put (300,149.4) {\line(1,0){20}}
\thinlines
\put (325,131.5) {$0^+$}
\put (325,144.4) {$2^+$}
\thicklines
\put (350,182.2) {\line(1,0){20}}
\thinlines
\put (375,177.2) {$0^+$}
\thicklines
\put (400,168.4) {\line(1,0){20}}
\put (400,204.3) {\line(1,0){20}}
\put (400,195.7) {\line(1,0){20}}
\put (400,207.7) {\line(1,0){20}}
\thinlines
\put (425,163.4) {$1^-$}
\put (425,197.3) {$2^-$}
\put (425,185.7) {$2^+$}
\put (425,209.7) {$3^+$}
\end{picture}
\vspace{15pt}
\caption[Energy levels of $^{12}$C]
{Comparison between the low-lying experimental spectrum 
of $^{12}$C \protect\cite{ajz} and that calculated with the 
oblate top Hamiltonian of Eq.~(\protect\ref{ham3}) with $N=10$. 
The parameter values are given in Table~\protect\ref{par}. 
States with uncertain spin-parity assignment are in parentheses.}
\label{ec12}
\end{figure}
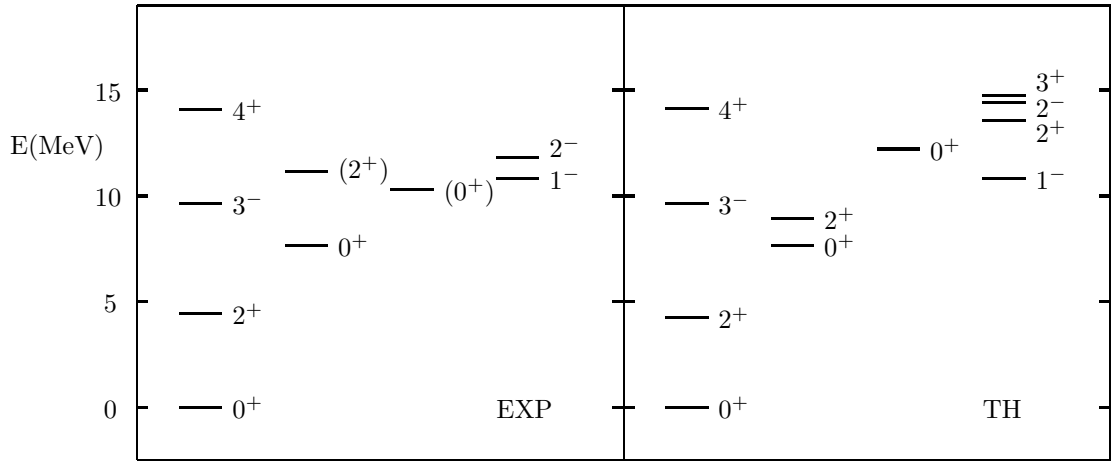

The Hamiltonian for the oblate top is given by Eq.~(\ref{ham3}). 
The coefficients $\xi_1$, $\xi_2$, $\kappa_1$ and $\kappa_2$ are 
determined in a fit to the excitation energies of $^{12}$C (see 
Table~\ref{par}). The number of bosons is taken to be $N=10$. 
In Fig.~\ref{ec12} we show a comparison between the experimental data 
and the calculated states of the oblate top with energies $< 15$ MeV.  
One can clearly identify in the experimental spectrum the states 
$0^{+}$, $2^{+}$, $3^{-}$, $4^{+}$ of the 
ground rotational band, the state $0^{+}$ of the stretching vibration 
and the state $1^{-}$ of the bending vibration. However, the $3^{-}$ state 
does not fall at the location expected from a rigid extended configuration 
with moments of inertia given by Eq.~(\ref{inertia}), but at a higher 
excitation energy. Since the $3^{-}$ state has $K=3$ the deviation from the 
simple rotational formula indicates large rotation-vibration interactions. 
This conclusion is strengthened by the low location of the stretching 
vibration. In contrast with molecules, the vibrational and rotational 
frequencies in nuclei appear to be comparable. If one takes as a 
measure of the ratio $E_{{\rm vib}}/E_{{\rm rot}}$ the quantity 
$E_{0_{2}^{+}}/E_{2_{1}^{+}}$, this quantity is $1.7$ in $^{12}$C and 
$\sim 10$ in a typical molecule with ${\cal D}_{3h}$ 
symmetry (for example, ozone O$_{3}$). 

\subsection{Form factors and transition rates}

Form factors for electron scattering on $^{12}$C have been measured long
ago. For the extended charge distribution of Eq.~(\ref{rhor}), the form 
factors are obtained by multiplying the oblate top form factors by an 
exponential factor $\exp (-q^{2}/4\alpha )$ (compare Eqs.~(\ref{fpoint}) 
and (\ref{fq})). The coefficient $\beta$ is determined from the first 
minimum in the elastic form factor at $q^2 = 3.4$ fm$^{-2}$ \cite{reuter} 
to be $\beta=1.74$ fm, and subsequently 
the coefficient $\alpha$ is determined from the charge radius of $^{12}$C,  
$\langle r^2 \rangle^{1/2} = 2.468 \pm 0.012$ fm \cite{reuter}, 
to be $\alpha=0.52$ fm$^{-2}$. If the alpha-particles would not be effected 
by the presence of the others, the coefficient $\alpha$ should be the same 
as for free particles. The value of $\alpha$ is slightly different from 
the one used in \cite{aurora}, where it was determined from the maximum 
in the elastic form factor. 

\begin{figure}

\vfill 

\begin{minipage}{.5\linewidth}
\centerline{\epsfig{file=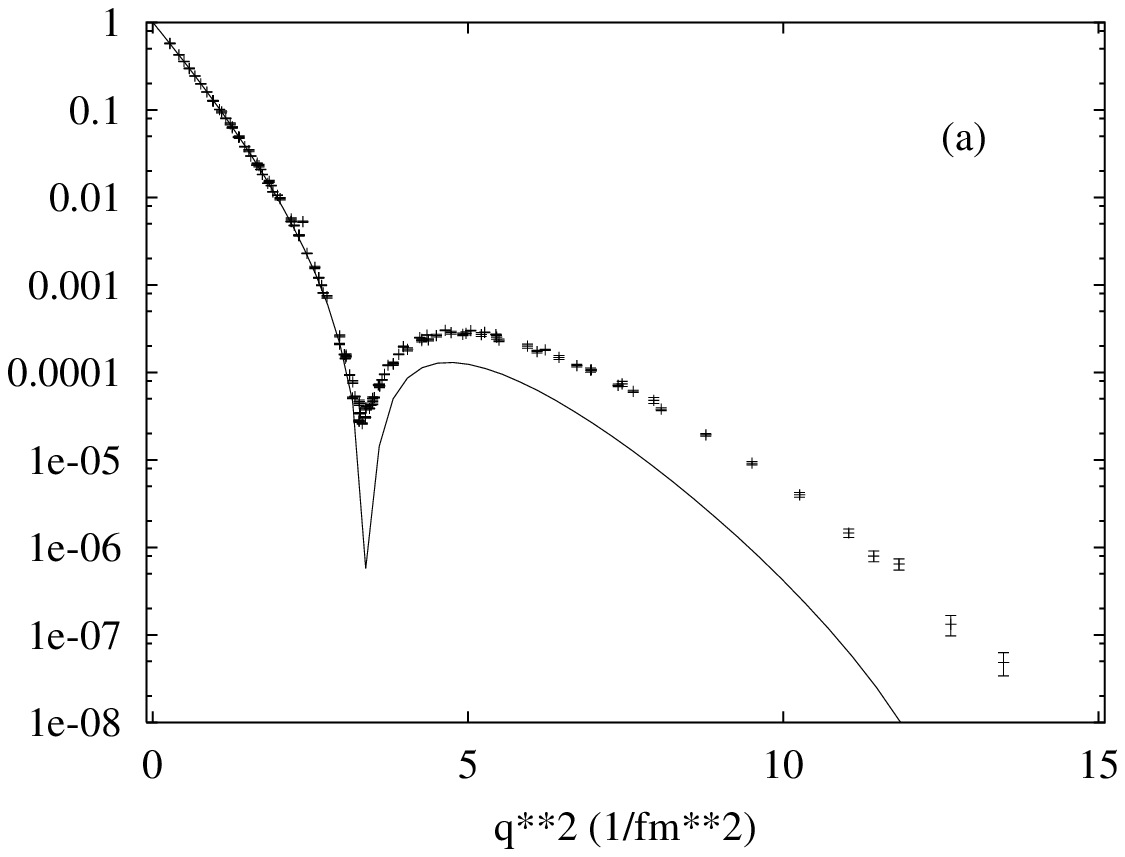,width=\linewidth}}
\end{minipage}\hfill
\begin{minipage}{.5\linewidth}
\centerline{\epsfig{file=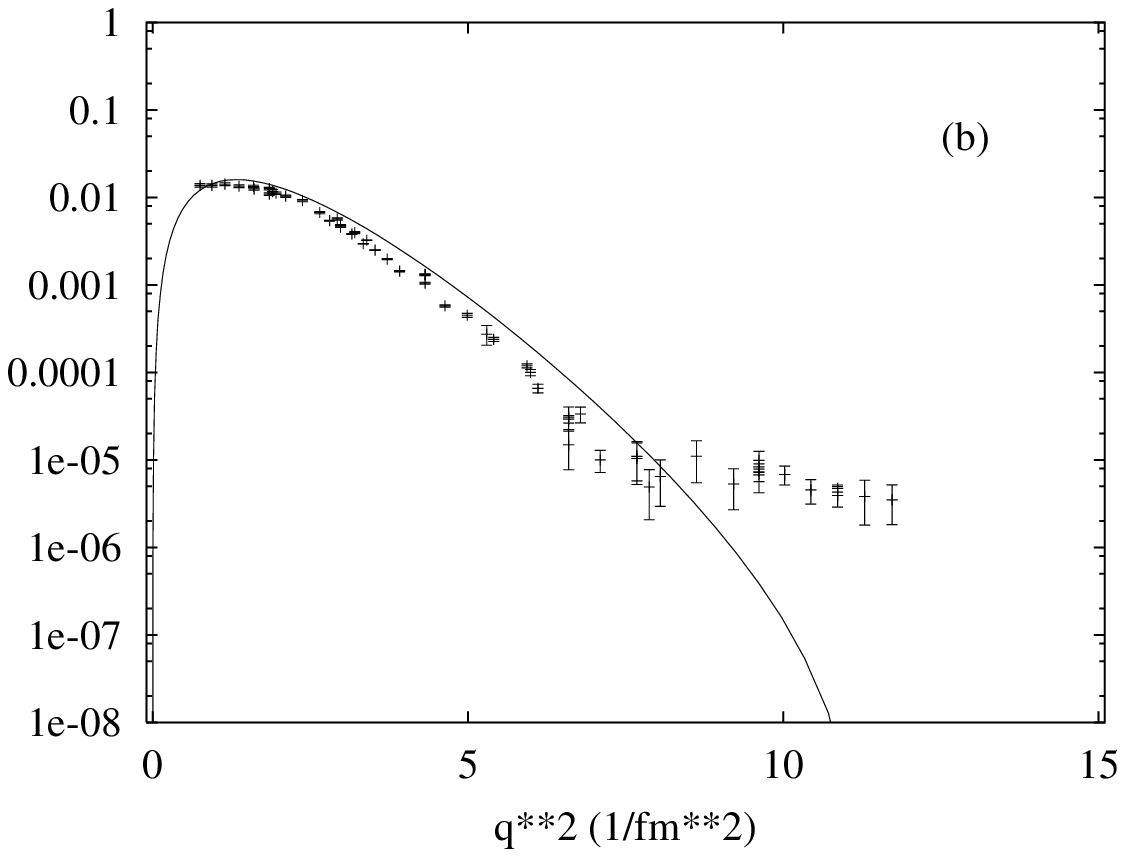,width=\linewidth}}
\end{minipage}

\begin{minipage}{.5\linewidth}
\centerline{\epsfig{file=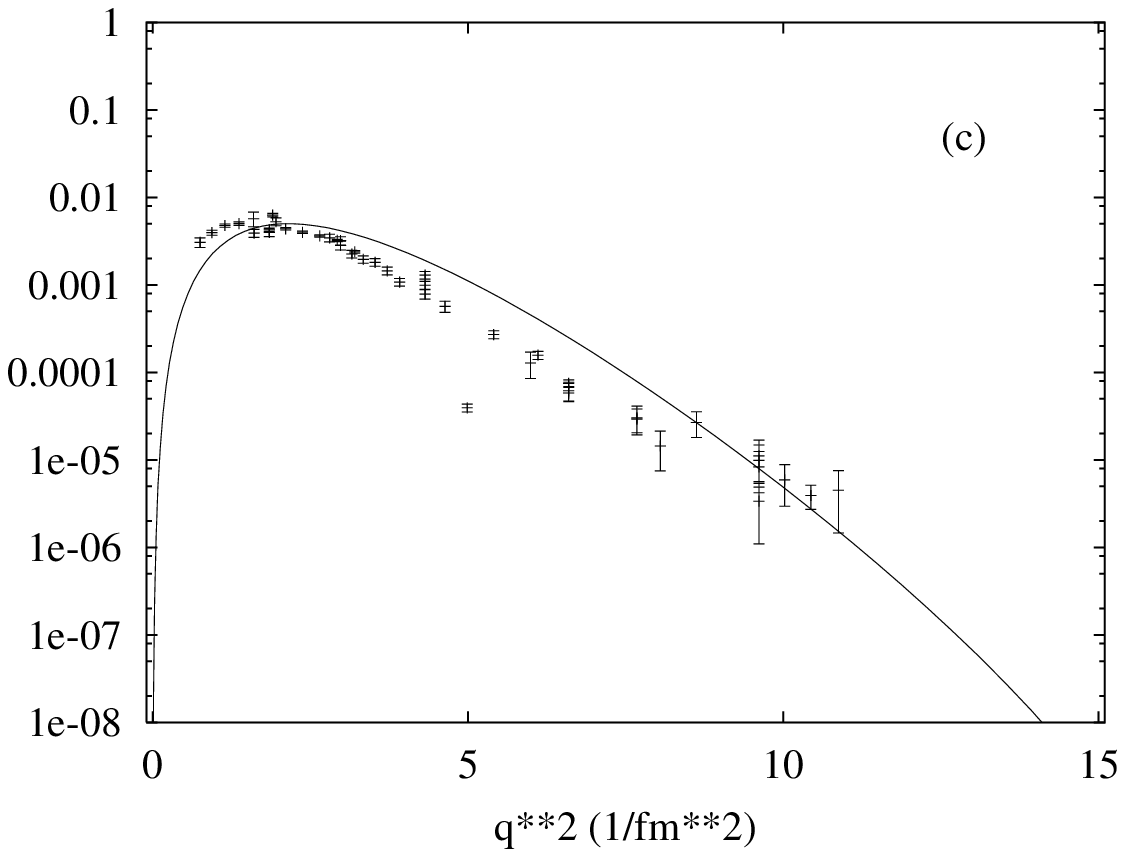,width=\linewidth}}
\end{minipage}\hfill
\begin{minipage}{.5\linewidth}
\centerline{\epsfig{file=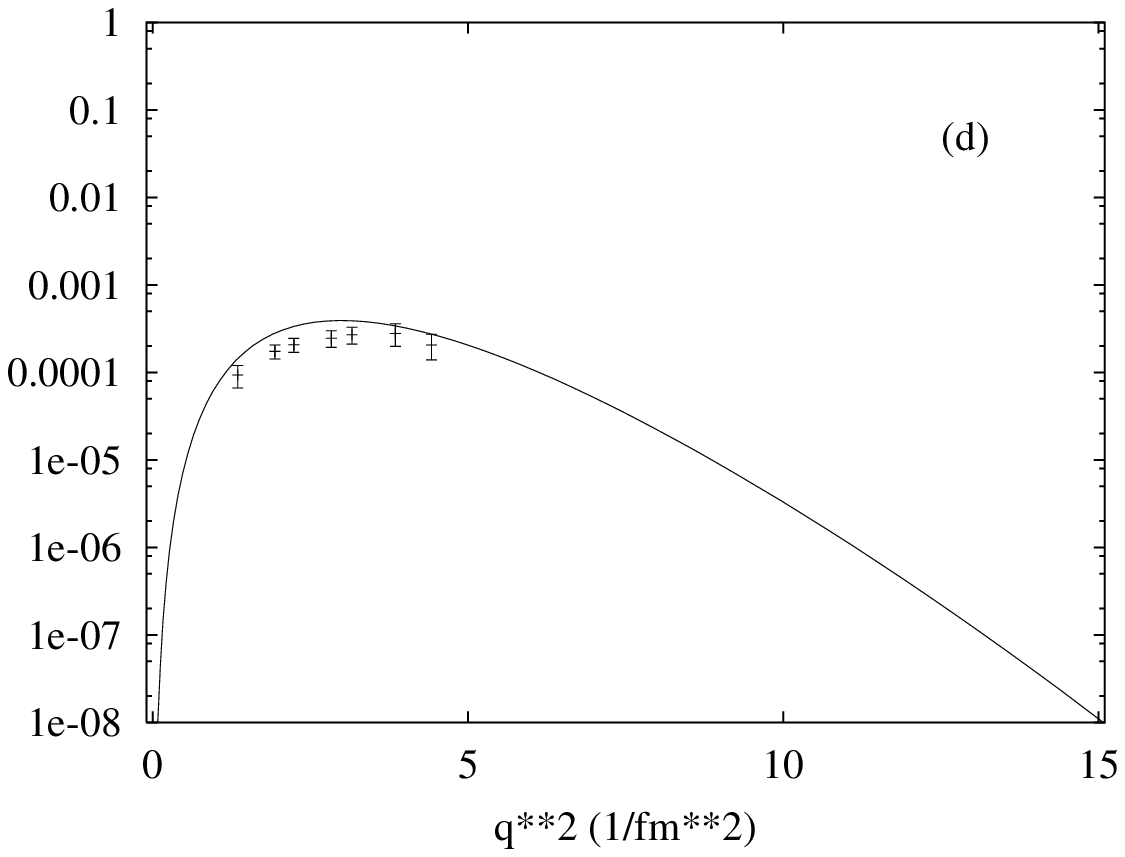,width=\linewidth}}
\end{minipage}

\begin{minipage}{.5\linewidth}
\centerline{\epsfig{file=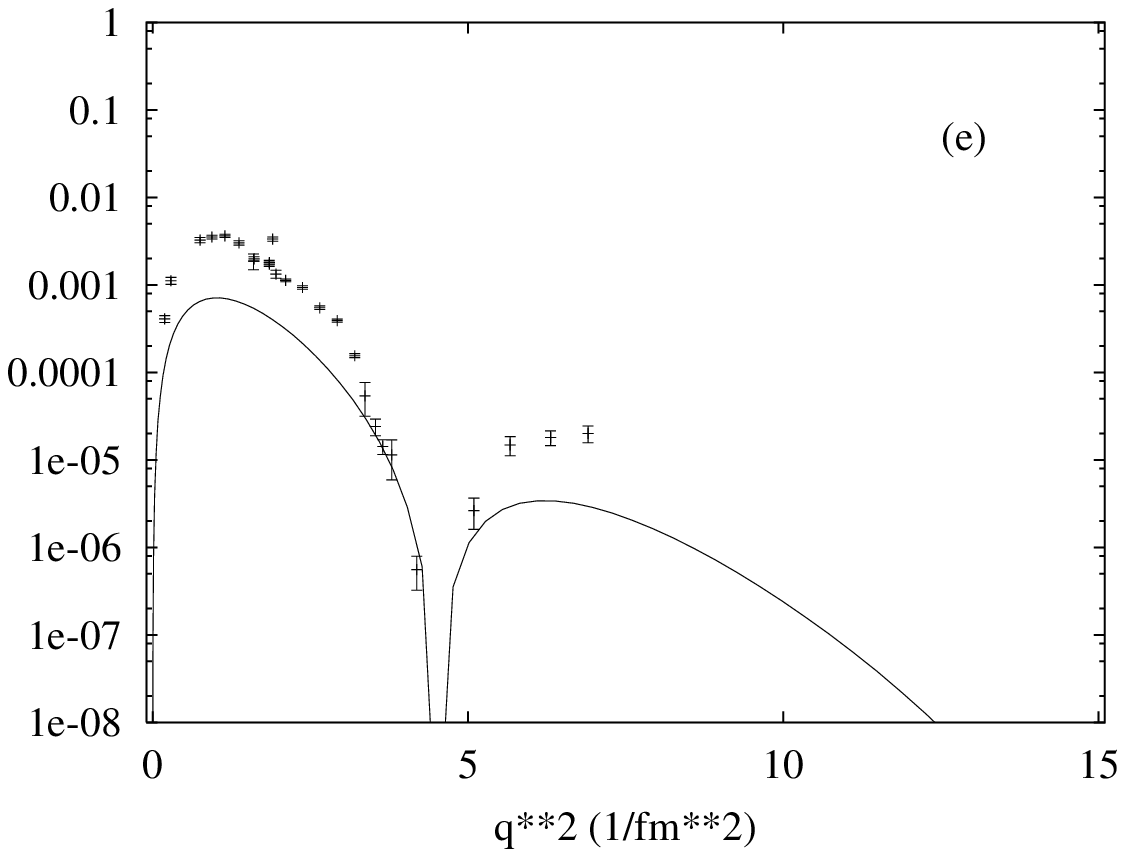,width=\linewidth}}
\end{minipage}\hfill
\begin{minipage}{.5\linewidth}
\centerline{\epsfig{file=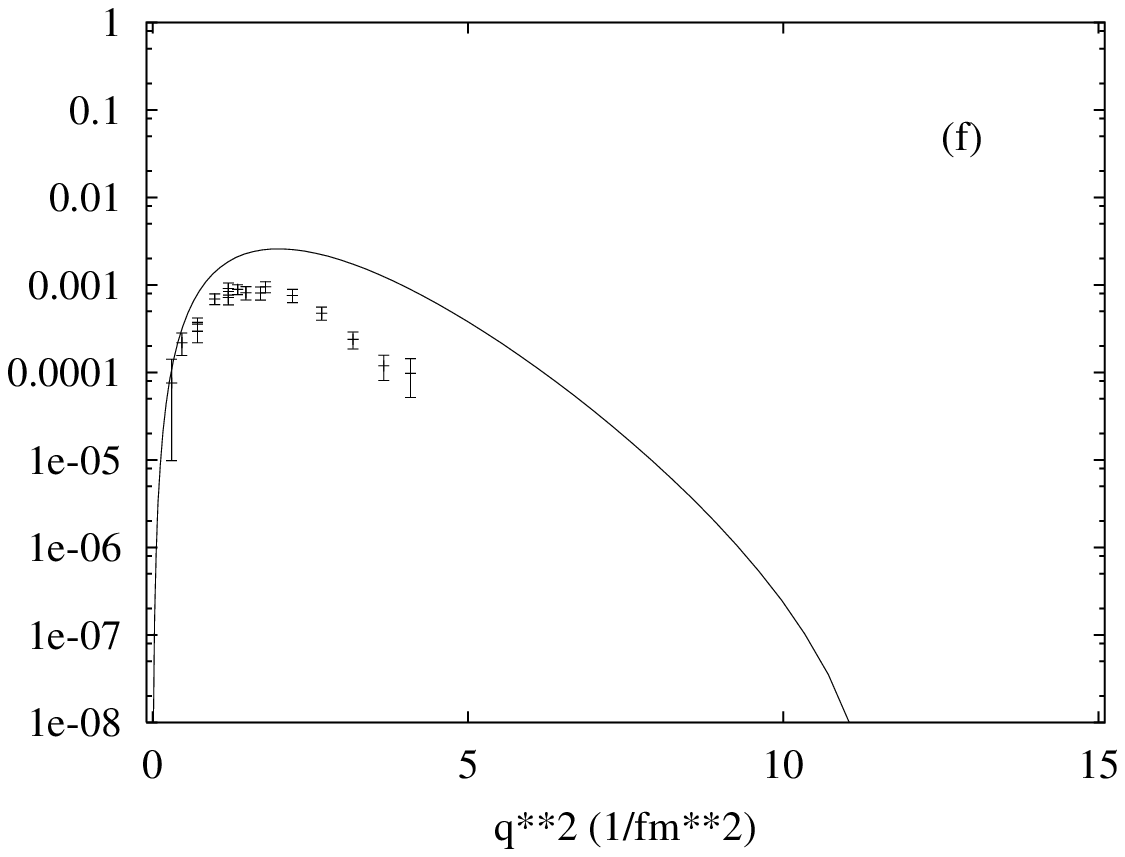,width=\linewidth}}
\end{minipage}

\caption[Form factors for $^{12}$C]
{Comparison between the experimental form factors 
$|{\cal F}(0^+_1 \rightarrow L^P_i;q)|^2$ of $^{12}$C 
for the final states (a) $L^P_i=0^+_1$ (elastic), (b) $L^P_i=2^+_1$, 
(c) $L^P_i=3^-_1$, (d) $L^P_i=4^+_1$, (e) $L^P_i=0^+_2$, and 
(f) $L^P_i=1^-_1$ and those obtained for the oblate top with $N=10$. 
The parameter values are given in Table~\protect\ref{par}. 
The experimental data are taken from \protect\cite{reuter} and 
\protect\cite{sick}-\protect\cite{nakada}.}
\label{ffc12}
\end{figure}

In Fig.~\ref{ffc12} we show a 
comparison between experimental and theoretical form factors calculated 
exactly using the Hamiltonian of Eq.~(\ref{ham3}). The value of the 
parameters of the Hamiltonian and the transition operator is given in 
Table~\ref{par}. We note that the transition form factors are calculated 
with the same values of coefficients $\alpha$ and $\beta$ as determined 
from the elastic form factor. Whereas for the rotational excitations the 
calculations seem to give a good description of the experimental data, 
the situation is different for vibrational excitations. 
Although the $q$ dependence of the 
vibrational form factors is consistent with experiment, indicating that the 
$0^{+}_2$ and $1^{-}_1$ excited states could indeed be the vibrational
excitations of a three-alpha configuration with ${\cal D}_{3h}$ symmetry,
their observed large strength implies a very large mixing with other
configurations. The strengths of these excitations are proportional 
to the corresponding intrinsic transition matrix elements $\chi_1$ and 
$\chi_2$ for each type of vibration $v_1$ and $v_2$, respectively \cite{BI}. 
According to Eq.~(\ref{chi}), the coefficients $\chi_{1}$ and $\chi_{2}$ 
go to zero in the large $N$ limit (rigid configurations). 
For finite values of $N$, the vibrational transition form factors are
different from zero. The solid lines in Figs.~\ref{ffc12} were obtained in 
calculations for a finite value of $N=10$. 
 
\begin{figure}
\centerline{\hbox{
\epsfig{figure=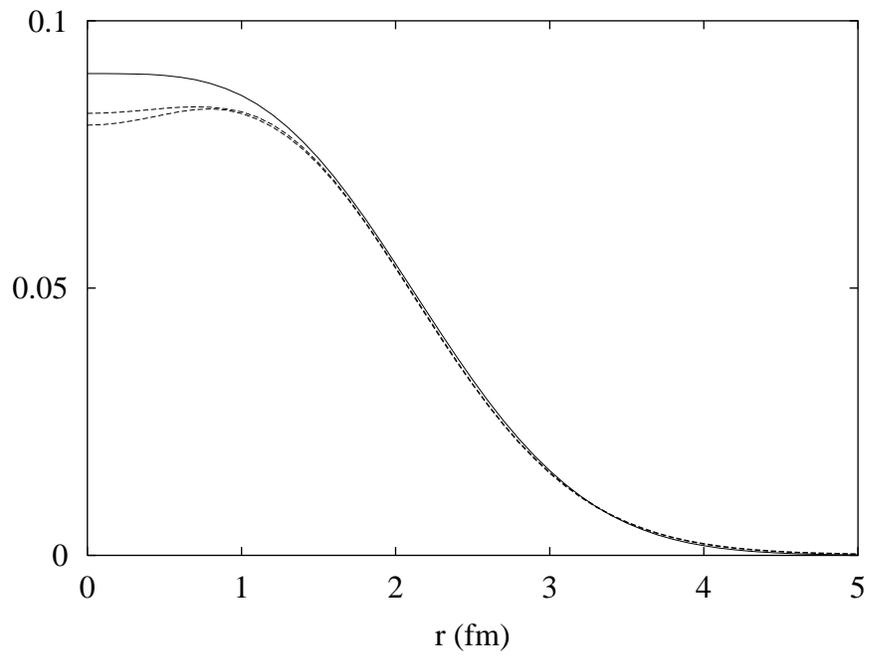} }}
\caption[Charge distribution $^{12}$C]
{Comparison between the ground state charge distribution of $^{12}$C 
extracted from the experimental elastic form factor (dashed lines) 
\protect\cite{reuter} and that calculated for the oblate top with 
$N=10$ (solid line). 
The parameter values are given in Table~\protect\ref{par}.} 
\label{gscharge}
\end{figure}

From the form factors one can extract other electromagnetic properties 
of interest. The ground state charge distribution can be obtained 
by taking the Fourier transform of the elastic form factor according to 
Eq.~(\ref{gsch}). In Fig.~\ref{gscharge} we compare the charge 
distribution of $^{12}$C obtained from the calculated elastic form 
factor of Fig.~\ref{ffc12} and extracted in a Fourier-Bessel analysis of 
the experimental data \cite{reuter}. 

The $B(EL)$ values correspond to the long wavelength limit of the form
factors (see Eq.~(\ref{belif})). The values extracted from the fit to the 
form factors are shown in Table~\ref{bem}. For the monopole strength, 
instead of $B(E0)$ it is
customary to give the matrix element $M(E0)$. While the $B(E2;2_{1}^{+}%
\rightarrow 0_{1}^{+})$ and $B(E3;3_{1}^{-}\rightarrow 0_{1}^{+})$ values
are in reasonable agreement, both the $B(E2;0_{2}^{+}\rightarrow 2_{1}^{+})$
and $M(E0;0_{2}^{+}\rightarrow 0_{1}^{+})$ deviate by an order of magnitude.
This again may indicate that the nature of the $0_{2}^{+}$ is not that of a
vibrational mode.

\section{Conclusions}

In this article, we have presented a detailed analysis of a configuration
composed of three alpha particles at the vertices of an equilateral
triangle. We have done so in a classical, quantum and quantum algebraic
description. The formulas that have been derived have been used to study \
the ground state configuration of $^{12}$C. In addition, we have introduced
a new method to study clustering based on the algebraic quantization of the
Jacobi variables. The latter method is very general and it allows to
describe in a relatively simple fashion not only three particles at the
vertices of a triangle (oblate top) but also other situations such as
six-dimensional vibrations and rotational-vibrational spectra.

Our analysis of the experimental data for $^{12}$C appears to indicate that
the triangular configuration describes the data (energy levels and form
factors) reasonably well for the rotational band $0_1^{+}$, $2_1^{+}$, 
$3_1^{-}$, $4_1^{+}$, although with large rotation-vibration interactions. 
The situation is different for the vibrational excitations $0_{2}^{+}$ 
and $1_{1}^{-}$. Here the shape of the form factors is well reproduced 
but its magnitude is not.

We bring attention to the quantum algebraic description in terms of the
algebra of $U(7)$ that is capable of describing not only rigid like
structures but also floppy structures. We point out that this method is very
general and can describe other situations, as, for example, three alpha
particles on a line. These configurations cannot describe the low-lying
experimental spectrum of $^{12}$C since for three alpha particles on a line
there are (in the ground rotational band) no negative parity states. They
may occur as excited configurations. Their properties will be presented in a
forthcoming publication. The crucial problem in clustering in nuclei is to
understand what type of configurations are present, if any, and to provide
unambiguous experimental evidence for these configurations. The algebraic
method gives a way in which all calculations can be performed easily.

Finally, the general algebraic framework based on the spectrum generating
algebra $U(7)$ also allows to describe clustering phenomena for three
different particles \cite{BL}. This description is relevant to giant trinuclear
molecules in ternary cold fission \cite{greiner,hess}.

\section*{Acknowledgements}

It is a pleasure to thank Aurora Tumino for her help 
in calculating the form factors of $^{12}$C. 
This work was supported in part by CONACyT under project 
32416-E, and by DPAGA under project IN106400, 
and by D.O.E. Grant DE-FG02-91ER40608.

\clearpage

\begin{table}
\centering
\caption[Transition form factors]
{Transition form factors ${\cal F}(0^+_1 \rightarrow L^P_i;q)$ 
for the harmonic oscillator (ho), the deformed oscillator (do) and the 
oblate top (ot). For each case, a scale parameter $\beta$ was introduced 
to give the same result for the charge radius.}
\label{ff}
\vspace{15pt}
\begin{tabular}{lccc}
\hline
& & & \\
$L^P_i$ & \multicolumn{3}{c}{${\cal F}(0^+_1 \rightarrow L^P_i;q)$} \\
& ho & do & ot \\
& & & \\
\hline
& & & \\
$0^+_1$ & $e^{-q^2 \beta^2/6}$ 
& $4 \, J_2(q\beta \sqrt{2})/q^2 \beta^2$ & $j_0(q \beta)$ \\
& & & \\
$0^+_2$ & $-\frac{1}{6\sqrt{3}} \, q^2 \beta^2 \, e^{-q^2 \beta^2/6}$ 
& $0$ & $- \chi_1 \, q \beta \, j_1(q \beta)$ \\
& & & \\
$1^-_1$ & $\frac{i}{6\sqrt{30}} \, q^3 \beta^3 \, e^{-q^2 \beta^2/6}$ 
& $i 8\sqrt{3} \, J_5(q\beta \sqrt{2})/q^2 \beta^2$ 
& $-i \chi_2 \, \frac{1}{2}\sqrt{3} \, q \beta \, j_2(q \beta)$ \\
& & & \\
$2^+_1$ & $-\frac{1}{3\sqrt{6}} \, q^2 \beta^2 \, e^{-q^2 \beta^2/6}$ 
& $-8\sqrt{2} \, J_4(q\beta \sqrt{2})/q^2 \beta^2$ 
& $\frac{1}{2} \sqrt{5} \, j_2(q \beta)$ \\
& & & \\
$3^-_1$ & $\frac{i}{18\sqrt{5}} \, q^3 \beta^3 \, e^{-q^2 \beta^2/6}$ 
& $i 8\sqrt{2} \, J_5(q\beta \sqrt{2})/q^2 \beta^2$ 
& $-i \sqrt{\frac{35}{8}} \, j_{3}(q\beta)$ \\
& & & \\
$4^+_1$ & $\frac{1}{18\sqrt{70}} \, q^4 \beta^4 \, e^{-q^2 \beta^2/6}$ 
& $\frac{48}{\sqrt{7}} \, J_6(q\beta \sqrt{2})/q^2 \beta^2$ 
& $\frac{9}{8}\,j_{4}(q\beta)$ \\
& & & \\
\hline 
\end{tabular}
\end{table}

%\clearpage

\begin{table}
\centering
\caption[Transition form factors: U(6) limit]
{Transition form factors ${\cal F}(0^+_1 \rightarrow L^P_i;q)$ 
for the $U(6)$ limit of the algebraic cluster model. The coefficient 
$\epsilon$ is given by $\epsilon=-q\beta/\sqrt{3N}$.}
\label{ffu6}
\vspace{15pt}
\begin{tabular}{lc}
\hline
& \\
$L^P_i$ & ${\cal F}(0^+_1 \rightarrow L^P_i;q)$ \\
& \\
\hline
& \\
$0^+_1$ & $(\cos \epsilon)^N$ \\
& \\
$0^+_2$ & $\sqrt{\frac{N!}{12(N-2)!}} \, (\cos \epsilon)^{N-2} 
\, (i \sin \epsilon)^2$ \\ 
& \\
$1^-_1$ & $\sqrt{\frac{N!}{40(N-3)!}} \, (\cos \epsilon)^{N-3} 
\, (i \sin \epsilon)^3$ \\
& \\
$2^+_1$ & $\sqrt{\frac{N!}{6(N-2)!}} \, (\cos \epsilon)^{N-2} 
\, (i \sin \epsilon)^2$ \\
& \\
$3^-_1$ & $\sqrt{\frac{N!}{60(N-3)!}} \, (\cos \epsilon)^{N-3} 
\, (i \sin \epsilon)^3$ \\
& \\
$4^+_1$ & $\sqrt{\frac{N!}{280(N-4)!}} \, (\cos \epsilon)^{N-4} 
\, (i \sin \epsilon)^4$ \\
& \\
\hline 
\end{tabular}
\end{table}

%\clearpage

\begin{table}
\centering
\caption[Transition form factors: SO(7) limit]
{Transition form factors ${\cal F}(0^+_1 \rightarrow L^P_i;q)$ 
for the $SO(7)$ limit of the algebraic cluster model. The coefficient 
$\epsilon$ is given by $\epsilon=-q\beta/\sqrt{N(N+5)/2}$.}
\label{ffso7}
\vspace{15pt}
\begin{tabular}{lc}
\hline
& \\
$L^P_i$ & ${\cal F}(0^+_1 \rightarrow L^P_i;q)$ \\
& \\
\hline
& \\
$0^+_1$ & $\frac{4!N!}{(N+4)!} \, C^{(5/2)}_N(\cos \epsilon)$ \\
& \\
$0^+_2$ & $0$ \\ 
& \\
$1^-_1$ & $\sqrt{\frac{189N!(N-3)!10!}{(N+4)!(N+7)!}}
\, (i\sin \epsilon)^3 \, C^{(11/2)}_{N-3}(\cos \epsilon)$ \\
& \\
$2^+_1$ & $\sqrt{\frac{140N!(N-2)!8!}{(N+4)!(N+6)!}}
\, (i \sin \epsilon)^2 \, C^{( 9/2)}_{N-2}(\cos \epsilon)$ \\
& \\
$3^-_1$ & $\sqrt{\frac{126N!(N-3)!10!}{(N+4)!(N+7)!}}
\, (i \sin \epsilon)^3 \, C^{(11/2)}_{N-3}(\cos \epsilon)$ \\
& \\
$4^+_1$ & $\sqrt{\frac{297N!(N-4)!12!}{(N+4)!(N+8)!}}
\, (i \sin \epsilon)^4 \, C^{(13/2)}_{N-4}(\cos \epsilon)$ \\
& \\
\hline 
\end{tabular}
\end{table}

%\clearpage

\begin{table}
\centering
\caption[Parameters]
{Coefficients used in the calculation of the energy spectrum and the 
transition form factors of $^{12}$C. The number of bosons is $N=10$.} 
\label{par}
\vspace{15pt}
\begin{tabular}{lrl}
\hline
& & \\
$\xi_1$    & 0.1721 & MeV \\
$\xi_2$    & 0.2745 & MeV \\
$\kappa_1$ & 0.7068 & MeV \\
$\kappa_2$ & 0.1276 & MeV \\
$R^2$      & 1.40 & \\
& & \\
$\beta$  & 1.74 & fm \\
$\alpha$ & 0.52 & fm$^{-2}$ \\
& & \\
\hline
\end{tabular}
\end{table}

%\clearpage

\begin{table}
\centering
\caption[$B(EL)$ values]{Comparison between calculated and 
measured $B(EL)$ values.}
\label{bem}
\vspace{15pt}
\begin{tabular}{cccll}
\hline
& & & & \\
& Th. & \multicolumn{2}{c}{Exp.} & Ref.\\
& & & & \\
\hline
& & & & \\
$B(E2;2_{1}^{+}\rightarrow 0_{1}^{+})$  & 8.4 
& $7.6 \pm 0.4$ & $e^{2}\mbox{fm}^{4}$ & \protect\cite{ajz} \\
$B(E3;3_{1}^{-}\rightarrow 0_{1}^{+})$ & 44 
& $103 \pm 17$ & $e^{2}\mbox{fm}^{6}$ & \protect\cite{ajz} \\
$B(E4;4_{1}^{+}\rightarrow 0_{1}^{+})$ & 73  
& & $e^{2}\mbox{fm}^{8}$ & \\  
$B(E2;0_{2}^{+}\rightarrow 2_{1}^{+})$ & 1.3  
& $13.1 \pm 1.8$ & $e^{2}\mbox{fm}^{4}$ & \protect\cite{ajz} \\
$M(E0;0_2^+ \rightarrow 0_1^+)$ & 0.4 
& $5.5 \pm 0.2$ & $\mbox{fm}^2$ & \protect\cite{strehl} \\
$\langle r^2 \rangle^{1/2}$ & 2.468  
& $2.468 \pm 0.12$ & fm & \protect\cite{reuter} \\ 
& & & \\
\hline
\end{tabular}
\end{table}

%\clearpage

\begin{figure}
\centering
\setlength{\unitlength}{1.0pt}
\begin{picture}(200,135)(0,0)
\thicklines
\put ( 25, 10) {\circle*{10}}
\put (125, 10) {\circle*{10}}
\put ( 75,110) {\circle*{10}}
\put ( 25, 10) {\line ( 1,0){100}}
\put ( 25, 10) {\line ( 1,2){ 50}}
\put (125, 10) {\line (-1,2){ 50}}
\put ( 25, 10) {\line ( 3, 2){ 50}}
\put (125, 10) {\line (-3, 2){ 50}}
\put ( 75,110) {\line ( 0,-1){ 67}}
\put ( 75,120) {1}
\put ( 10,  0) {2}
\put (135,  0) {3}
\put (150, 85) {\vector(0,1){25}}
\put (150, 85) {\vector(1,0){25}}
\put (185, 80) {$\hat x$}
\put (145,120) {$\hat z$}
\end{picture}
\caption[Geometric configuration]{Geometry of a three-body system.}
\label{geometry}
\end{figure}
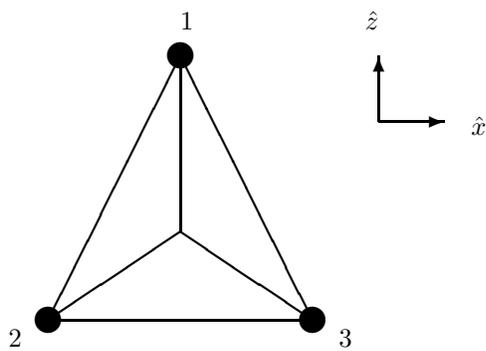

%\clearpage

\begin{figure}
\centering
\setlength{\unitlength}{1.0pt} 
\begin{picture}(260,320)(0,0)
\thinlines
\put (  0,  0) {\line(1,0){260}}
\put (  0,320) {\line(1,0){260}}
\put (  0,  0) {\line(0,1){320}}
\put (260,  0) {\line(0,1){320}}
\thicklines
\put ( 50, 30) {\line(1,0){20}}
\put ( 50,140) {\line(1,0){20}}
\put ( 50,195) {\line(1,0){20}}
\put ( 50,250) {\line(1,0){20}}
\put (140,140) {\line(1,0){20}}
\put (140,250) {\line(1,0){20}}
\put (190,250) {\line(1,0){20}}
%\multiput ( 60,30)(0,5){45}{\circle*{0.1}}
\thinlines
\put ( 20,280) {$n \backslash \sigma$}
\put ( 60,280) {$n$}
\put ( 30, 25) {$0$}
\put ( 30,135) {$2$}
\put ( 30,190) {$3$}
\put ( 30,245) {$4$}
\put ( 75, 25) {$0^+$}
\put ( 75,135) {$2^+$}
\put ( 75,190) {$3^-,\,1^-$}
\put ( 75,245) {$4^+,\,2^+,\,0^+$}
\put (140,280) {$n-2$}
\put (165,135) {$0^+$}
\put (165,245) {$2^+$}
\put (190,280) {$n-4$}
\put (215,245) {$0^+$}
\put (150, 25) {$U(6) \supset SO(6)$}
\end{picture}
\vspace{15pt}
\caption[Spectrum: U(6) limit] 
{Schematic spectrum of the harmonic oscillator with $U(6) \supset
SO(6)$ symmetry. The number of bosons is $N=4$. All states are symmetric
under $S_3$.}
\label{hosc}
\end{figure}

%\clearpage

\begin{figure}
\centering
\setlength{\unitlength}{1.0pt} 
\begin{picture}(260,340)(0,-20)
\thinlines
\put (  0,-20) {\line(1,0){260}}
\put (  0,320) {\line(1,0){260}}
\put (  0,-20) {\line(0,1){340}}
\put (260,-20) {\line(0,1){340}}
\thicklines
\put ( 40, 30) {\line(1,0){20}}
\put ( 40,102) {\line(1,0){20}}
\put ( 40,156) {\line(1,0){20}}
\put ( 40,222) {\line(1,0){20}}
\put (135,162) {\line(1,0){20}}
\put (135,234) {\line(1,0){20}}
\put (190,246) {\line(1,0){20}}
\multiput ( 90, 30)(5,0){25}{\circle*{0.1}}
\multiput (147.5, 30)(0,5){27}{\circle*{0.1}}
\multiput (200, 30)(0,5){44}{\circle*{0.1}}
\thinlines
\put ( 20,280) {$\sigma \backslash \omega$}
\put ( 50,280) {$4$}
\put ( 30, 25) {$0$}
\put ( 30, 97) {$2$}
\put ( 30,151) {$3$}
\put ( 30,217) {$4$}
\put ( 65, 25) {$0^+$}
\put ( 65, 97) {$2^+$}
\put ( 65,151) {$3^-,\,1^-$}
\put ( 65,217) {$4^+,\,2^+,\,0^+$}
\put (145,280) {$2$}
\put (125,157) {$0$}
\put (125,229) {$2$}
\put (160,157) {$0^+$}
\put (160,229) {$2^+$}
\put (200,280) {$0$}
\put (180,241) {$0$}
\put (215,241) {$0^+$}
\put (150,  5) {$SO(7) \supset SO(6)$}
\end{picture}
\vspace{15pt}
\caption[Spectrum: SO(7) limit]
{Schematic spectrum of a deformed oscillator with $SO(7)$ symmetry.
The number of bosons is $N=4$. All states are symmetric under $S_3$.}
\label{defosc}
\end{figure}
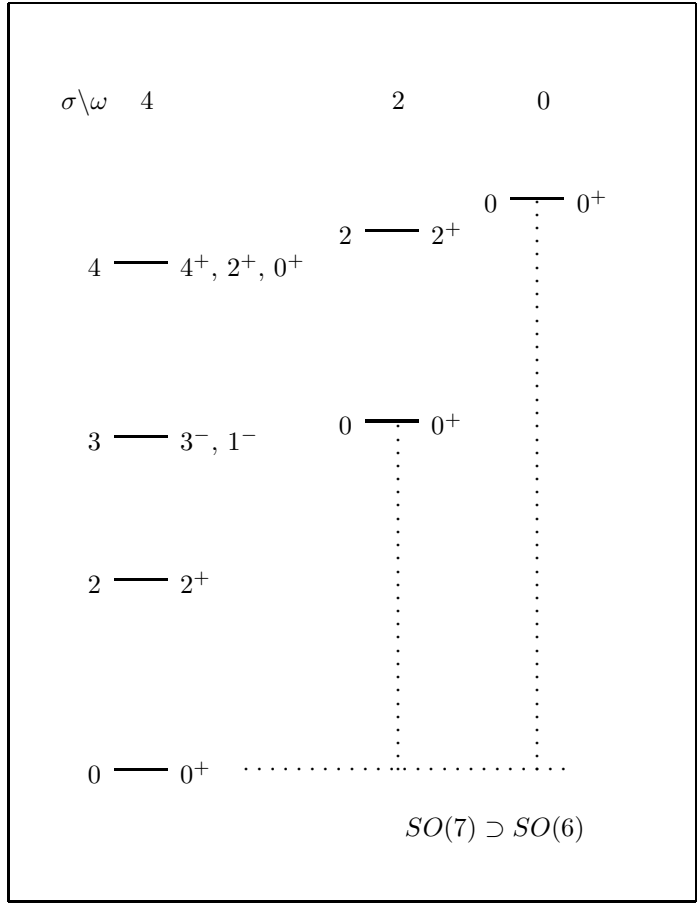

%\clearpage

%\clearpage

%\clearpage

%\clearpage

\clearpage

\clearpage

%\clearpage

\end{document}